\documentclass{article}
\usepackage[english]{babel}

\usepackage[a4paper,top=2cm,bottom=2cm,left=3cm,right=3cm,marginparwidth=1.75cm]{geometry}

\usepackage{amsmath}
\usepackage{amsfonts}  % or \usepackage{amssymb}

\usepackage{graphicx}
\usepackage[colorlinks=true, allcolors=blue]{hyperref}
\usepackage{booktabs}
\usepackage{pdflscape} 

\title{Robust MCVaR Portfolio Optimization with Ellipsoidal Support and Reproducing Kernel Hilbert Space-based Uncertainty}
\author{
  Rupendra Yadav\thanks{Department of Mathematics, Indian Institute of Technology Delhi, Hauz Khas, New Delhi 110016, India. Email: \texttt{rupendra.yadav@maths.iitd.ac.in}. Corresponding author.}
  \and
  Aparna Mehra\thanks{Department of Mathematics, Indian Institute of Technology Delhi, Hauz Khas, New Delhi 110016, India. Email: \texttt{apmehra@maths.iitd.ac.in}.}
}

\date{}
\begin{document}
\maketitle

\begin{abstract}
This study introduces a portfolio optimization framework to minimize mixed conditional value at risk (MCVaR), incorporating a chance constraint on expected returns and limiting the number of assets via cardinality constraints. A robust MCVaR model is presented, which presumes ellipsoidal support for random returns without assuming any distribution. The model utilizes an uncertainty set grounded in a reproducing kernel Hilbert space (RKHS) to manage the chance constraint, resulting in a simplified second-order cone programming (SOCP) formulation. The performance of the robust model is tested on datasets from six distinct financial markets. The outcomes of comprehensive experiments indicate that the robust model surpasses the nominal model, market portfolio, and equal-weight portfolio with higher expected returns, lower risk metrics, enhanced reward-risk ratios, and a better value of Jensen's alpha in many cases.
Furthermore, we aim to validate the robust models in different market phases (bullish, bearish, and neutral). The robust model shows a distinct advantage in bear markets, providing better risk protection against adverse conditions. In contrast, its performance in bullish and neutral phases is somewhat similar to that of the nominal model. The robust model appears effective in volatile markets, although further research is necessary to comprehend its performance across different market conditions.
\vspace{1em}

\noindent\textbf{Keywords:} Portfolio Optimization, Cardinality Constraints, Conditional Value-at-Risk, Robust Optimization, Reproducing Kernel Hilbert Space
\end{abstract}

\section{Introduction}
The pioneering work of Markowitz \cite{markowitz1952portfolio}  established the foundational principles of modern portfolio theory (MPT), providing a mathematical framework for optimizing portfolios and balancing risk against expected returns. Markowitz advocated a diversification strategy to mitigate portfolio risk, demonstrating that combining assets with low correlations could reduce overall portfolio volatility. His seminal work marked the beginning of an era in financial economics where the relationship between risk and return became central to investment theory.
The following decades witnessed various scholars expand upon and refine the MPT framework. Sharpe’s \cite{sharpe1964capital} development of the Capital Asset Pricing Model (CAPM) further solidified the link between risk, return, and asset pricing. In parallel, new risk and performance measures were proposed to address the limitations of earlier models. These include the Sharpe ratio, which measures risk-adjusted return, and the Sortino ratio, which focuses on downside risk \cite{sharpe1966mutual,sortino1994performance}.

Other influential risk metrics include the conditional value-at-risk (CVaR) and value-at-risk (VaR), which offer a more comprehensive understanding of risk by focusing on tail risk and extreme market events \cite{jorion1996value}.  VaR is a quantile risk measure at a defined significance level, whereas CVaR is the expected value of all losses that surpass VaR, assessing the investment's risk level. CVaR is a coherent risk measure \cite{artzner1997thinking}, exhibiting increased sensitivity to the tail and shape of the return distribution, so capturing extreme losses more proficiently than VaR \cite{ghaoui2003worst}.  Utilizing sampling theory, the portfolio optimization problem that minimizes CVaR is formulated as a linear program, providing computing benefits \cite{rockafellar2000optimization}.

%  \textcolor{red}{Building on this approach, a CVaR optimization model with proportional costs to generate an optimal derivative portfolio is proposed in \cite{alexander2003derivative}.
% In \cite{yu2009portfolio}, Lévy processes and the Variance Gamma Copula (VGC) capture the skewness and kurtosis of asset returns to minimize the CVaR of portfolio loss. A multi-stage portfolio selection model incorporating conditional value-at-risk (CVaR), solved using a hybrid genetic algorithm and particle swarm optimization, with parameter tuning via Taguchi experimental design in \cite{najafi2015multi}. In \cite{ban2018machine}, a performance-based regularization (PBR) approach is used to constrain the sample variances of the estimated portfolio risk and return for the mean-conditional value-at-risk portfolio optimization problem. 
%  }
The paper \cite{nasini2022multi} demonstrates that the multi-market portfolio optimization problem with CVaR constraints is NP-hard and proposes a decomposition strategy combined with strong valid inequalities to solve the problem efficiently. An AI-powered technique for stock price prediction via filtering and CVaR optimization is presented in \cite{wang2024two}. 

%Beyond the realm of finance, CVaR has numerous applications across various fields, including power engineering, mobile edge computing, blockchain, and others \cite{you2023cvar}, \cite{asensio2015stochastic}, \cite{ge2024research}, \cite{cui2023portfolio}, \cite{xu2023towards}.}
To enable a more intricate representation of risk by assessing many risk levels simultaneously rather than focusing on a single quantile, In  \cite{mansini2007conditional},
the mixed conditional value at risk (MCVaR) for a tolerance set \( \{ \delta_1,  \dots, \delta_m \} \), $0\leq \delta_m< \delta_{m-1}<\ldots <\delta_1< 1$, is defined as the weighted sum of the individual CVaR values at different levels of significance. The formal expression is given by
\[
\text{MCVaR}_{\{\delta_1, \delta_2, \dots, \delta_m \}} = \sum_{j=1}^m \theta_j  \text{CVaR}_{\delta_j}\,,
\]
where $\theta_j \in (0,1),\; j=1,\ldots, m,$ are weights.
This MCVaR enhances risk understanding by including the complete characteristics of the return distribution and considering multiple thresholds. The model's ability to apply varying weights to different quantiles allows it to prioritize risk levels, making it especially advantageous in contexts where specific tail occurrences or certain risk thresholds are of greater importance. In \cite{goel2018index}, MCVaR is used to design portfolios for index tracking (IT) problems. 

The omega ratio, among others, provides an alternative perspective on performance evaluation by incorporating higher-order moments of the return distribution, thereby providing a more comprehensive assessment of risk-adjusted performance compared to the Sharpe ratio \cite{keating2002universal}. Investors also seek to evaluate the performance of their portfolios relative to a benchmark index or to compare different portfolio strategies. Performance metrics like the Treynor ratio \cite{treynor1965rate} and Jensen alpha \cite{jensen1968performance} are essential financial tools for conducting such assessments.

In addition to advancing risk metrics, researchers have highlighted the importance of asset management for optimal portfolio outcomes in uncertain markets. Among the constraints imposed in the classical mean-variance optimization model, cardinality constraints—which limit the asset count in the portfolio—are vital \cite{chang2000heuristics}. These constraints simplify handling portfolios, enhance asset quality, and reduce transaction cost \cite{khodier2024adapted}. It aids liquidity and aligns with preferences and ESG standards \cite{kraussl2024review}. Solving mean-CVaR optimization with cardinality constraints is challenging due to mixed integer formulation \cite{bertsimas2022scalable}. Kobayashi et al. \cite{kobayashi2021bilevel} designed a bilevel cutting-plane algorithm for exactly solving such problems.

Traditional mean-risk portfolio optimization frameworks enforce hard constraints on the requisite minimum return threshold while minimizing portfolio risk. To introduce flexibility in the anticipated return, chance-constrained optimization incorporates market uncertainty into the optimization model by dictating the constraint be satisfied with a specified probability instead of absolute certainty \cite{krokhmal2002portfolio}. Formally, a chance constraint can be written as 
\[
P\left(g(x, \xi\right) \leq 0) \geq 1 - \epsilon,
\]
where \( g(x, \xi) \) represents the constraint function that depends on both decision variables \( x \) and the random uncertain parameters \( \xi \) (such as returns) and \( \epsilon>0\) is a small probability that indicates the acceptable risk in $g(x,\xi)\leq0$ constraint violation.

Calafiore and El-Ghaoui \cite{calafiore2006distributionally} studied the robust optimization approach for chance-constrained linear programs for a class of radially symmetric probability distributions. K\"{u}\c{c}\"{u}kyavuz and Jiang \cite{jiange2022} presented a survey on chance-constrained problems when only limited information on the distribution or the moments of the distribution is available. While chance-constrained approaches offer a probabilistic framework, they assume that the distributions of returns are known and accurately modelled \cite{bonami2009exact}. However, in practice, these distributions may be misspecified, or there may be significant model uncertainty. This can lead to suboptimal decisions if the true risk deviates from the assumed probability distributions \cite{chen2020sparse}.

Robust optimization seeks to find solutions that perform well under a range of possible scenarios, known as candidate distributions, rather than relying on precise estimates of parameters \cite{ben2000robust,bertsimas2004price}.
Parametric uncertainty can also arise from the underlying probability distributions that describe the returns of the assets. These distributions represent the statistical models that assume specific characteristics, such as means, variances, and correlations, but these models may not fully capture the true behaviour of the returns \cite{delage2010distributionally}. 

The worst-case analysis of CVaR under mixed, box, and ellipsoid
uncertainty sets of the underlying probability distribution is proposed in \cite{zhu2009worst}. In \cite{ling2012robust}, a robust portfolio selection model
under combined marginal and joint ellipsoidal uncertainty set is proposed. The worst-case CVaR minimization by considering the most adverse scenarios in the return vectors is studied in \cite{gotoh2013robust}. 
In \cite{kara2019stability}, a robust CVaR optimization model is proposed using the construction of a parallelepiped ambiguity set. In \cite{gu2024data}, under different support sets (ellipsoidal, polytopic, and unbounded), tractable conic reformulations of robust two-stage stochastic programming with the mean-CVaR criterion are established.

In \cite{yang2022kernel}, the kernel ambiguity set is constructed using Kernel Mean Embedding (KME) and Maximum Mean Discrepancy (MMD). The primary contribution of \cite{yang2022kernel} lies in developing a novel method based on the kernel ambiguity set, which offers distinct advantages over existing distributionally robust optimization (DRO) approaches that rely on moment-based information. 

In our work, we apply the DRO method to handle chance constraints on portfolio expected returns, which leverage the kernel ambiguity set to address uncertainty in returns.
A comparative study of portfolio optimization techniques is given in \cite{gunjan2023brief}, where the authors compiled classical, machine learning-based, and quantum-inspired techniques.

The multi-fold contributions of the present paper are summarized as:
\begin{itemize}
    \item \textbf{Formulation of the MCVaR minimization problem:}
    We put forward a portfolio optimization model that endeavours to minimize a mixed conditional value-at-risk (MCVaR), subject to constraints on portfolio expected returns, asset investment limits, and restrictions on asset count within the portfolio. Additionally, we formulated a robust model of posit ellipsoidal support for uncertain returns without imposing any distribution assumption.
      
    \item \textbf{Chance constraint-based model and RKHS-based uncertainty set:} We establish a probabilistic chance constraint on the expected return from a portfolio. We utilize an ambiguity set grounded in Reproducing Kernel Hilbert Space (RKHS) to convert this constraint into a more manageable form. The design of this kernel-based ambiguity set includes the use of Maximum Mean Discrepancy (MMD), which serves to measure the disparity between the uncertain return rates and the sampled distribution. The MMD is reformulated using kernel techniques to produce a tractable Second-Order Cone Programming (SOCP) optimization problem.

    \item \textbf{Performance evaluation across six market datasets:} We perform extensive numerical evaluations of the proposed models using the rolling window approach. Our tests utilize six datasets: Nikkei 225 (Japan), S\&P 100 (USA), NIFTY 50 (India), FTSE 100 (UK), Dow Jones Industrial Average (DJIA) (USA), and BOVESPA (Brazil). We assess the risk-return performance through 13 statistical metrics that include return, risk, risk-adjusted ratios, and Jensen's alpha. The findings reveal that (a) across all reviewed markets, the robust models reliably achieve higher average returns than the nominal model, with significant outperformance observed in particular scenarios such as \( A = 6 \) within the DJIA index and \( A = 15 \) in the BOVESPA index, where \( A \) represents the count of invested assets.
    (b) The robust models demonstrate lower VaR and CVaR values in most scenarios while maintaining competitive standard deviations, particularly in the NIFTY 50 and FTSE 100 indices.
    (c) The robust models achieve superior reward-to-risk ratios (e.g., Sharpe, Sortino, Treynor, and STARR ratios) and favourable Jensen’s Alpha across various market conditions, with outstanding performance for \( A = 9 \) in most indices.

    \item \textbf{Market phase analysis:}
    We analyze the performance of the models in three market phases: the bull phase (positive market trend), the bear phase (negative market trend), and the neutral phase. We observe that the proposed robust model with three assets performed significantly better in a bearish market, yielding nearly half the negative return of its nominal counterpart. However, the robust and nominal performance is comparable during bullish and neutral market phases. Further research is needed to draw a definitive conclusion about the model's efficacy on varying market phases.

\end{itemize}

The rest of this paper is organized as follows:  Section \eqref{section:2} presents the problem formulation and the development of the nominal MCVaR minimization model. We put forward a chance-constrained MCVaR minimization model. Section \eqref{section:3} presents reformulations of chance constraint under ellipsoidal support and RKHS-based uncertainty. Section \eqref{section:4} describes the experimental setup, including the market datasets and performance measures. Section \eqref{section:5} discusses the out-of-sample analysis of the proposed models and performance comparison with some other portfolios. We conclude with directions for future research in Section \eqref{section:6}.

\section{Mixed Conditional Value-at-Risk Portfolio Optimization}\label{section:2}
Let \( w \) represent a portfolio composed of \( n \) assets, denoted as \( w = ( w_1, \dots, w_n )^{'} \), where \( w_i \) is a decision variable indicating the proportion of the total capital allocated to the \( i \)-th asset, for \( i = 1,  \dots, n \). The investment horizon is typically divided into \( T \) discrete periods to observe the return realizations for each asset. Let \( r_{ij} \) denote the return of the \( i \)-th asset at \( j \)-th time, with the probability of this realization being \( p_j \). The expected return for an asset \( i \) is given by:
\[
\mu_i = \sum_{j=1}^{T} r_{ij} p_j, \quad i = 1,  \dots, n.
\]

The portfolio \( w \) return at the \( j \)-th time, denoted by \( R_j(w) \), is the weighted sum of the asset returns:
\[
R_j(w) = \sum_{i=1}^{n} r_{ij} w_i, \quad \text{with probability} \ p_j, \quad j = 1,  \dots, T.
\]

The portfolio return \( R(w) \) is thus a finite random variable $ \left( R_1(w), \dots, R_T(w) \right)$
accompanied by the probability vector \( \left( p_1,  \dots, p_T \right) \). It is worth noting that these models equally apply to scenarios with arbitrary distributions or sets of realizations.

The expected return \( \mu(R_w) \) of the portfolio \( w \) can be approximated as:
\[
\mu(R_w) \approx\sum_{j=1}^{T} p_j R_j(w) =\sum_{j=1}^{T} \sum_{i=1}^{n} p_j r_{ij} w_i.
\]

The return \( r_{ij} \) for the \( i \)-th asset in the \( j \)-th scenario is computed as:
\[
r_{ij} = \frac{C_{i,j} - C_{i,j-1}}{C_{i,j-1}}\,,
\]
where \( C_{i,j} \) is the closing price of the \( i \)-th asset at \( j \)-th time. It is assumed that short selling is prohibited, and the scenarios are considered equally likely by setting \( p_j = 1/T \), for \( j = 1,  \dots, T \).

The maximum investment in the \(i\)-th asset is constrained by:
\[
w_i \leq u_i y_i, \quad i = 1,  \dots, n,
\]
where,
\(u_i\) is the upper bound on investment, and variable \( y_i \) is a binary decision variable that takes the value $1$ if the \( i \)-th asset is included in the portfolio and $0$ otherwise.
Similarly, the minimum investment in the \(i\)-th asset is bounded by:
\[
w_i \geq l_i y_i, \quad i = 1,  \dots, n,
\]
where, \(l_i\) is the lower bound on investment in the \(i\)-th asset. The vector $y \text{ is defined as } \left(y_1, \ldots, y_n\right)$. The number of assets invested in the portfolio is given by constraint $\sum_{i=1}^{n} y_i = A$,  where an investor decides $A$.

The mixed conditional value-at-risk (MCVaR) is a weighted combination of multiple CVaR risk measures calculated at different confidence levels \( \delta \). For \( m \) distinct values of \( \delta_k,\; k = 1, \dots, m, \) and \( 0 \leq \delta_m < \dots < \delta_1 < 1 \), the MCVaR is defined as:

\[
\text{MCVaR}(R(w)) = \theta_1 \text{CVaR}_{\delta_1}(R(w)) + \dots + \theta_m \text{CVaR}_{\delta_m}(R(w)),
\]

where \( \theta_k > 0,\;  k = 1, \dots, m \), are weights satisfying the condition \( \sum_{k=1}^{m} \theta_k = 1 \).

%The set \( \Theta \) is defined as:
%\[
%\Theta = \Big\{ \theta = (\theta_1, \dots, %\theta_m)' : \theta_k \geq 0,\; k = 1, %\dots, m, \; \sum_{k=1}^{m} \theta_k = 1 %\Big\}.
%\]
The mean-MCVaR$(R(w))$ optimization problem with the above constraints is given as follows:
\begin{align}
\min_{w, y} \quad & \text{MCVaR$(R(w))$} \nonumber\\
\text{subject to}\quad & \sum_{i=1}^{n} w_{i} =1, \nonumber\\
%& 0 \leq w_{i} \leq 1,\quad  i=1, \ldots,n \\
&l_{i} y_{i}\leq w_{i} \leq u_{i} y_{i}, \quad i = 1, \ldots, n, \nonumber\\
%& w_{i} \geq , \quad i = 1, \ldots, n \\
& y_i \in \{0,1\}, \quad i = 1, \ldots, n, \nonumber \\
& \sum_{i=1}^{n} y_i = A, \nonumber\\
& \mu(R_w) \geq R_{\ast}, \nonumber
\end{align}
where $R_{\ast}$ is the minimum desired return level. 

Analogous to the minimization of CVaR at a given confidence level $\gamma$ that can be approximated by a linear programming problem when dealing with continuous distributions \cite{rockafellar2000optimization}, the minimization of MCVaR$(R(w))$ can be written as \cite{mansini2007conditional}:
\begin{align}
\text{(NoM)} \quad  \min_{c, y, w} \quad & \sum_{k=1}^{m} \theta_k \Big( \gamma_k + \frac{1}{\delta_k} \sum_{j=1}^{T} c_{jk} p_j \Big) \nonumber\\
\text{subject to} \quad & c_{jk} + \gamma_k + \sum_{i=1}^{n} r_{ij} w_i \geq 0, \quad k = 1, \ldots, m, \; j = 1, \ldots, T,\nonumber \\
& c_{jk} \geq 0, \quad k = 1, \ldots, m, \; j = 1, \ldots, T, \nonumber \\
& \sum_{i=1}^{n} w_{i} =1, \nonumber\\
%& 0 \leq w_{i} \leq 1,\quad i=1, \ldots,n, \nonumber \\
& l_{i} y_{i} \leq w_{i} \leq u_{i} y_{i}, \quad i = 1, \ldots, n, \nonumber \\
& y_i \in \{0,1\}, \quad i = 1, \ldots, n,  \nonumber\\
& \sum_{i=1}^{n} y_i = A, \nonumber\\
& \mu(R_w) \geq R_{\ast}, \nonumber
\end{align}
where,
$
c_{jk} = \left(- \sum_{i=1}^{n} r_{ij} w_i - \gamma_k\right)^{+}, \; k = 1, \ldots, m, \; j = 1, \ldots T,$ are auxiliary variables,  and $R_{\ast}$ is the minimum level of expected return desired from the investment. We shall refer to the above model as the nominal model and the optimal portfolio obtained from it by NoMP.

\subsubsection*{Chance Constraint Model}
While the nominal model provides a useful framework, it does not consider the uncertainty often present in the financial market. We extend the (NoM) model by imposing a chance constraint on the expected return at the violation level of $\Gamma$.

The general formulation of the MCVaR minimization equipped with the chance constraint on expected return is given as follows:
\begin{align}
\min_{w, y} \quad & \text{MCVaR$(R(w))$} \nonumber\\
\text{subject to}\quad & \sum_{i=1}^{n} w_{i} =1, \nonumber\\
%& 0 \leq w_{i} \leq 1,\quad  i=1, \ldots,n \\
&l_{i} y_{i}\leq w_{i} \leq u_{i} y_{i}, \quad i = 1, \ldots, n, \nonumber\\
%& w_{i} \geq , \quad i = 1, \ldots, n \\
& y_i \in \{0,1\}, \quad i = 1, \ldots, n, \nonumber \\
& \sum_{i=1}^{n} y_i = A, \nonumber\\
& P \{ \mu(R_w) \geq R_{\ast}\} \geq 1 - \Gamma, \nonumber
\end{align}
where $\Gamma \in (0, 1)$  is the risk tolerance in achieving the expected return higher than $R_{\ast}$ from the portfolio.

The above model is equivalently reformulated as follows:
\begin{align}
\text{(CCM)} \quad  \min_{c, y, w} \quad &\sum_{k=1}^{m} \theta_k \Big( \gamma_k + \frac{1}{\delta_k} \sum_{j=1}^{T} c_{jk} p_j \Big)\nonumber \\
\text{subject to} \quad & c_{jk} + \gamma_k + \sum_{i=1}^{n} \tilde{r}_{ij} w_i \geq 0, \quad k = 1, \ldots, m, \; j = 1, \ldots, T, \nonumber \\
& c_{jk} \geq 0, \quad k = 1, \ldots, m, \; j = 1, \ldots, T, \nonumber \\
& \sum_{i=1}^{n} w_{i} =1,\nonumber \\
&  l_{i} y_{i}\leq w_{i} \leq u_{i} y_{i}, \quad  i = 1, \ldots, n, \nonumber \\
& y_i \in \{0,1\}, \quad  i = 1, \ldots, n,\nonumber \\
& \sum_{i=1}^{n} y_i = A, \nonumber\\
& P \{ \mu(R_w) \geq R_{\ast} \} \geq 1 - \Gamma.\label{eq:chance}
\end{align}

%Here $ c_{jk} = \left(- \sum_{i=1}^{n} \tilde{r}_{ij} w_i - \gamma_k\right)^{+}, \quad k = 1, \ldots, m, \; j = 1, \ldots T,$ are auxiliary variables.

\section{Proposed Robust MCVaR  Model}\label{section:3}
Let $(S, \mathcal{F})$ be a measurable space with $\mathcal{F}$ being the Borel $\sigma$-algebra on $S$.
The random return vector $\tilde{r}_{j} \in \mathbb{R}^n$ is defined on a probability space $(S, \mathcal{F}, P)$ with a closed support set $\mathcal{V}_j$, and $P$ represents the associated probability measure of $\tilde{r}_{j}$.

\subsection{Reformulation under Ellipsoidal Support}
We assume that uncertain return rates ($\tilde{r}_j$) belong to a family of sets defined as  
\[
\mathcal{V}_j = \{r_j + P_j v \,:\, v \in \mathbb{R}^n, \; \|v\|_2 \leq 1 \}, \quad j = 1, \ldots, T,
\]
where \( r_j \in \mathbb{R}^n\) is a return realization at time $j,\;j= 1, \ldots, T$, \( P_j \in \mathbb{R}^{n \times n} \) is a full-rank matrix defining a linear transformation. 

Each set \( \mathcal{V}_j \) is an ellipsoid in \( \mathbb{R}^n \), commonly referred to as an ellipsoidal support set. The vector \( r_j \) defines the centre of the ellipsoid \( \mathcal{V}_j \), while the matrix \( P_j \) determines its shape and orientation. The principal axes of the ellipsoid are aligned with the columns of \( P_j \), and their lengths are proportional to the singular values of \( P_j \).

The worst-case representation of the constraints 
\[
-c_{jk} - \gamma_k - \sum_{i=1}^{n} \tilde{r}_{ij} w_i \leq 0, \quad k = 1, \ldots, m, \; j = 1, \ldots, T,
\]  
can be formulated in the problem (CCM) as follows:
\begin{align}
\min_{c, y, w} \quad & \sum_{k=1}^{m} \theta_k \Big( \gamma_k + \frac{1}{\delta_k} \sum_{j=1}^{T} c_{jk} p_j \Big) \nonumber\\
\text{subject to} \quad & \max_{\tilde{r}_j \in \mathcal{V}_j} \Big(-c_{jk} - \gamma_k - \sum_{i=1}^{n} \tilde{r}_{ij} w_i \Big) \leq 0, \quad k = 1, \ldots, m, \; j = 1, \ldots, T, \nonumber \\
& c_{jk} \geq 0, \quad k = 1, \ldots, m, \; j = 1, \ldots, T, \nonumber\\
& \sum_{i=1}^{n} w_i = 1, \nonumber\\
& l_i y_i \leq w_i \leq u_i y_i, \quad i = 1, \ldots, n, \nonumber\\
& y_i \in \{0, 1\}, \quad i = 1, \ldots, n, \nonumber\\
& \sum_{i=1}^{n} y_i = A, \nonumber\\
& P \{ \mu(R_w) \geq R_\ast \} \geq 1 - \Gamma. \nonumber
\end{align}

The inner maximization in the constraint  
\[
\max_{\tilde{r}_j \in \mathcal{V}_j } \Big(-c_{jk} - \gamma_k - \sum_{i=1}^{n} \tilde{r}_{ij} w_i \Big),
\]  
can be simplified using the definition of the ellipsoidal set \( \mathcal{V}_j \). Substituting \( \tilde{r}_j = r_j + P_j v \) into the expression, we obtain:  
\[
\max_{\tilde{r}_j \in \mathcal{V}_j} \Big(-c_{jk} - \gamma_k - \sum_{i=1}^{n} \tilde{r}_{ij} w_i \Big) 
= \max_{\|v\|_2 \leq 1} \Big(-c_{jk} - \gamma_k - \sum_{i=1}^{n} (r_{ij} + P_j v) w_i\Big).
\]  
Simplifying this, the term can be expressed as:
\[
-c_{jk} - \gamma_k - \sum_{i=1}^{n} r_{ij} w_i + \max_{\|v\|_2 \leq 1 } ( -v^{'} P_j^{'} w ).
\]  
Using the dual norm property, \(  \max_{\|v\|_2 \leq 1 }( -v^{'} P_j^{'} w )\) is equivalent to the Euclidean norm \( \|P_j^{'} w\|_2 \). Thus, the constraint reduces to:  
\[
-c_{jk} - \gamma_k - \sum_{i=1}^{n} r_{ij} w_i + \|P_j^{'} w\|_2 \leq 0, \quad k = 1, \ldots, m, \; j = 1, \ldots, T.
\]
The problem (CCM) finally becomes:
\begin{align}
\text{(CCM-1)} \quad \min_{c, y, w} \quad &\sum_{k=1}^{m} \theta_k \Big( \gamma_k + \frac{1}{\delta_k} \sum_{j=1}^{T} c_{jk} p_j \Big) \nonumber\\
\text{subject to} \quad & -c_{jk} - \gamma_k - \sum_{i=1}^{n} r_{ij} w_i + \|P_j^{'} w\|_{2}\leq 0, \quad  k = 1, \ldots, m, \; j = 1, \ldots, T,  \nonumber \\
& c_{jk} \geq 0, \quad  k = 1, \ldots, m, \;  j = 1, \ldots, T,  \nonumber\\
& \sum_{i=1}^{n} w_{i} =1,  \nonumber \\
%& 0 \leq w_{i} \leq 1,\quad \forall i=1, \ldots,n \\
& l_{i} y_{i}\leq w_{i} \leq u_{i} y_{i}, \quad  i = 1, \ldots, n,  \nonumber \\
& y_i \in \{0,1\}, \quad  i = 1, \ldots, n,  \nonumber\\
& \sum_{i=1}^{n} y_i = A,  \nonumber\\
& P \{ \mu(R_w) \geq R_{\ast} \} \geq 1 - \Gamma.  \nonumber
\end{align}

\subsection{Kernel Reformulation of Chance Constraint}

Let \( \mathcal{P} \) represent the set of all probability measures in the measurable space \( (S, \mathcal{F}) \). We define \( \mathbf{H} \) as a Hilbert space of real-valued functions and let \( \beta: \mathcal{V}_j \to \mathbf{H} \) be a map such that the reproducing property holds for all \( h \in \mathbf{H} \) and \( \tilde{r}_j \in \mathcal{V}_j \). Denote \( \nu \) as the kernel mean embedding of distribution \( P \), and let \( \nu_{P_N} \) represent the kernel mean embedding of the empirical distribution.

We define the Hilbert space norm ball of radius \( \alpha \) as:
\[
\mathcal{B} = \left\{ \nu \in \mathbf{H} : \| \nu - \nu_{P_N} \|_{\mathbf{H}} \leq \alpha \right\},
\]
where \( \| \nu - \nu_{P_N} \|_{\mathbf{H}} \) denotes the maximum mean discrepancy (MMD). Building upon the work of Yang et al. \cite{yang2022kernel}, we consider the kernel mean embedding-based ambiguity set for the probability distribution of \( \tilde{r}_j \), given by,
\[
\mathcal{U}_{\mathcal{B}} = \left\{ P \in \mathcal{P} :
P(\tilde{r}_j \in \mathcal{V}_j) = 1, \;
\int \beta \, dP = \nu, \, \nu \in \mathcal{B} \subseteq \mathbf{H}
\right\}.
\]
The chance constraint in (CCM-1)
can be expressed as:
\begin{equation}
P ( R_{\ast}-\mu(R_w) \leq 0) \geq 1 - \Gamma.  \nonumber
\end{equation}
Thus, the worst-case reformulation under the ambiguity set $\mathcal{U}_{\mathcal{B}}$ is given by:
\begin{equation}
\min_{P(\tilde{r}) \in \mathcal{U}_{\mathcal{B}}} \, P (R_{\ast}-\mu(R_w) \leq 0 ) \geq 1 - \Gamma.  \nonumber
\end{equation}
Which can be rewritten as:
\begin{equation}
\max_{P(\tilde{r}) \in \mathcal{U}_{\mathcal{B}}} P( R_{\ast}-\mu(R_w) \geq 0 ) \leq \Gamma.  \nonumber
\end{equation}
The CVaR approximation of this constraint is then formulated as:
\begin{equation}
    \max _{P(\tilde{r}) \in \mathcal{U}_{\mathcal{B}}} \operatorname{CVaR}_{\Gamma}(R_{\ast}-\mu(R_w)) \leq 0. \label{cvar}
\end{equation}
Furthermore, CVaR can be approximated by the following optimization problem:
\[
\min _{\lambda \in \mathbb{R}}\Big(\lambda+\frac{1}{\Gamma} \mathbb{E}_{P(\tilde{r})}\Big(R_{\ast}-\mu(R_w)-\lambda\Big)^{+}\Big) \leq 0. \nonumber
\]
Therefore, the expression in \eqref{cvar} can be reformulated as: 
\[
\max_{P(\tilde{r}) \in \mathcal{U}_{\mathcal{B}}}\min _{\lambda \in \mathbb{R}}\Big(\lambda+\frac{1}{\Gamma} \mathbb{E}_{P(\tilde{r})}\Big(R_{\ast}-\mu(R_w)-\lambda\Big)^{+}\Big) \leq 0.  \nonumber
\]
If the ambiguity set $\mathcal{U}_{\mathcal{B}}$ is a compact convex set, then $\max$ and $\min$ can be interchanged, i.~e. 
\begin{equation}
\min _{\lambda \in \mathbb{R}}\Big(\lambda+\frac{1}{\Gamma} \max _{P(\tilde{r}) \in \mathcal{U}_{\mathcal{B}}} \mathbb{E}_{P(\tilde{r})}\Big(R_{\ast}-\mu(R_w)-\lambda\Big)^{+}\Big)\leq 0.\label{innermin}
\end{equation}
The inner optimization problem with constraints from the ambiguity set $\mathcal{U}_\mathcal{B}$ can be written as:
\begin{align}
\max _{P(\tilde{r}) \in \mathcal{U}_{\mathcal{B}}}\quad& \mathbb{E}_{P(\tilde{r})}\Big(R_{\ast}-\mu(R_w)-\lambda\Big)^{+} \nonumber\\
\text{subject to} \quad & \left\| \nu - \nu_{P_N} \right\|_{\textbf{H}} \leq \alpha, \nonumber\\
& \int \beta(\tilde{r}) dP(\tilde{r}) = \nu.  \nonumber
\end{align}
By using the empirical embedding, \( \nu \) and \( \nu_{P_N} \) can be written as 
\[
\nu = \sum_{j=1}^{T} \eta_j \beta(r_j) \quad \text{and} \quad \nu_{P_N} = \sum_{j=1}^{T_0} \frac{1}{T_0} \beta(r_j),
\]
where \( \{ r_j \}_{j=1}^{T} \) are expansion vectors used to discretize the support of the candidate distributions. Specifically,  \( \{ r_j \}_{j=1}^{T_0} \) are the sample return rates, and \( \{r_j \}_{j=T_0+1}^{T} \) are the support vectors. The probability mass on \(r_j\) is denoted by \(\eta_j\).

With the empirical embedding of \(\nu\) and \(\nu_{P_N}\), the above optimization problem can be reformulated as:
\begin{align}
\max _{\eta}\quad & \sum_{j=1}^{T} \eta_{j} \Big(R_{\ast}- \sum_{i=1}^{n} r_{ij} w_i-\lambda\Big)^{+}  \nonumber\\
\text {subject to}\quad & \left\|\sum_{j=1}^{T} \eta_{j} \beta\left(r_j\right)-\sum_{j=1}^{T_0} \frac{1}{T_0} \beta\left(r_j\right)\right\|_{\textbf{H}} \leq \alpha, \label{eq:alpha}\\
& \sum_{j=1}^{T} \eta_{j}=1,  \nonumber \\
& \eta_{j} \geq 0, \quad  j=1, \ldots, T.   \nonumber
\end{align}

Using the kernel function \(K(\cdot, \cdot)\) associated with \textbf{H}, $MMD^2$ can be reformulated as:  
\[
MMD^2 = \mathbb{E}_{r_l, r_k \sim P} \left[K(r_l, r_k)\right] - 2\mathbb{E}_{r_l \sim P, r_k \sim P_0} \left[K(r_l, r_k)\right] + \mathbb{E}_{r_l, r_k \sim P_0} \left[K(r_l, r_k)\right],
\]
where \(P\) and \(P_N\) denote the distributions from which the samples \(\{r_j\}_{j=1}^T\) and \(\{r_j\}_{j=1}^{T_0}\), respectively, are drawn. 

Given empirical samples, the empirical estimate of \(MMD^2\) is computed as:  
\[
\widehat{MMD}^2 = \sum_{l=1}^{T} \sum_{k=1}^{T} \eta_l \eta_k K(r_l, r_k) - \frac{2}{T_0} \sum_{l=1}^{T} \sum_{k=1}^{T_0} \eta_l K(r_l, r_k) + \frac{1}{T_0^2} \sum_{l=1}^{T_0} \sum_{k=1}^{T_0} K(r_l, r_k).
\]
As a result, the constraint \eqref{eq:alpha} can be reformulated as:
\begin{equation}
\sum_{l=1}^{T}\sum_{k=1}^{T}\eta_l \eta_k K(r_l,r_k)- \frac{2}{T_0} \sum_{l=1}^{T}\sum_{k=1}^{T_{0}}\eta_l K(r_l,r_k)+\frac{1}{T_0^{2}} \sum_{l=1}^{T_{0}}\sum_{k=1}^{T_{0}}K(r_l,r_k) \leq \alpha^2.  \nonumber
\end{equation}
Let, $R(w)=\left(\left(R_{\ast}- \sum_{i=1}^{n} r_{i1} w_i-\lambda\right)^{+}, \ldots,\left(R_{\ast}- \sum_{i=1}^{n} r_{iT} w_i-\lambda\right)^{+}\right)^{'}$. Then, the optimization problem can be reformulated as a quadratic-constrained linear objective optimization problem:
\begin{align}
\max _{\eta}\quad & R(w)^{'} \eta  \nonumber\\
\text {subject to} \quad & \sum_{l=1}^{T}\sum_{k=1}^{T}\eta_l \eta_k K(r_l,r_k)- \frac{2}{T_0} \sum_{l=1}^{T}\sum_{k=1}^{T_{0}}\eta_l K(r_l,r_k)+\frac{1}{T_0^{2}} \sum_{l=1}^{T_{0}}\sum_{k=1}^{T_{0}}K(r_l,r_k) \leq \alpha^2,  \nonumber \\
& \sum_{j=1}^{T} \eta_{j}=1,  \nonumber \\
& \eta_{j} \geq 0, \quad  j=1, \ldots, T. \nonumber 
\end{align}

The above problem can finally be reformulated as a second-order cone programming (SOCP) problem as follows:
\begin{align}
\max _{\eta}\quad & R(w)^{'} \eta  \nonumber \\
\text {subject to} \quad& {\left[\begin{array}{c}
L \eta \\
\frac{1}{T_0} \sum_{l=1}^{T}\sum_{k=1}^{T_{0}}\eta_l K(r_l,r_k) \\
\frac{1}{T_0} \sum_{l=1}^{T}\sum_{k=1}^{T_{0}}\eta_l K(r_l,r_k)
\end{array}\right]-\left[\begin{array}{c}
\textbf{0} \\
-\frac{\alpha^2}{2}+\frac{1}{2}+\frac{1}{2 T_0^{2}} \sum_{l=1}^{T_{0}}\sum_{k=1}^{T_{0}}K(r_l,r_k) \\
-\frac{\alpha^2}{2}-\frac{1}{2}+\frac{1}{2 T_0^{2}} \sum_{l=1}^{T_{0}}\sum_{k=1}^{T_{0}}K(r_l,r_k)
\end{array}\right] \in \mathfrak{C}^{T+2}},  \nonumber \\
& \sum_{j=1}^{T} \eta_{j}=1,  \nonumber \\
& \eta_{j} \geq 0, \quad  j=1, \ldots, T,  \nonumber
\end{align}
where $\mathfrak{C}^{T+2}$ denotes the $(T+2)$-dimension quadratic cone, \textbf{0} is a column zero vector, and $L$ is the Cholesky decomposition of the Gram matrix associated with kernel $K$ i.~e., $LL^{'} = \{K(r_l,r_k)\}_{l=1, k=1}^{l=T, k=T}$ and $M = \{K(r_l,r_k)\}_{l=1, k=1}^{l=T, k=T_0}$.

Using the conic duality, the dual formulation of the above SOCP is given as follows:
\begin{align}
\min_{\omega,\beta_1,\beta_2,\Phi} \quad &-\omega-\left(1-\alpha^2+\frac{1}{T_0^{2}} \sum_{l=1}^{T_{0}}\sum_{k=1}^{T_{0}}K(r_l,r_k)\right) \frac{\beta_{1}}{2}+\left(1+\alpha^2-\frac{1}{T_0^{2}} \sum_{l=1}^{T_{0}}\sum_{k=1}^{T_{0}}K(r_l,r_k)\right) \frac{\beta_{2}}{2} \nonumber\\
\text {subject to}\quad & \omega\textbf{1} +L^{'} \Phi+\frac{\beta_{1}}{T_{0}} M \textbf{1} +\frac{\beta_{2}}{T_{0}} M \textbf{1} +I\zeta =-
R(w), \label{eq:zeta}\\
& \|\Phi\|_{2}^{2}+\beta_{1}^{2} \leq \beta_{2}^{2}, \nonumber \\
&\omega,\,\beta_1,\,\beta_2\in \mathbb{R}, \;\Phi,\, \zeta \in \mathbb{R}^T,\; \zeta_{j} \geq 0, \quad  j=1, \ldots, T, \nonumber
\end{align}
where \textbf{1} is a column vector of ones of appropriate dimension. 

Since $\zeta\geq 0$, the constraint \eqref{eq:zeta} can be made more tractable by converting it to an inequality constraint. Thus,
 the above model can be rewritten as:

\begin{align}
\min_{\omega,\beta_1,\beta_2,\Phi} \quad &-\omega-\left(1-\alpha^2+\frac{1}{T_0^{2}} \sum_{l=1}^{T_{0}}\sum_{k=1}^{T_{0}}K(r_l,r_k)\right) \frac{\beta_{1}}{2}+\left(1+\alpha^2-\frac{1}{T_0^{2}} \sum_{l=1}^{T_{0}}\sum_{k=1}^{T_{0}}K(r_l,r_k)\right) \frac{\beta_{2}}{2} \label{ex1}  \\
\text{subject to}\quad & \omega\textbf{1} +L^{'} \Phi+\frac{\beta_{1}}{T_{0}} M \textbf{1} +\frac{\beta_{2}}{T_{0}} M \textbf{1}  \leq-
R(w),\label{ex2} \\
& \|\Phi\|_{2}^{2}+\beta_{1}^{2} \leq \beta_{2}^{2}.\label{ex3}
\end{align} 
By replacing $\max _{P(\tilde{r}) \in \mathcal{U}_{\mathcal{B}}} \mathbb{E}_{P(\tilde{r})}\left(R_{\ast}-\mu(R_w)-\lambda\right)^{+}$ with expressions \eqref{ex1}-\eqref{ex3}, the minimization problem \eqref{innermin} changes to
\begin{align}
& \min_{\lambda \in \mathbb{R}} \Big\{ \lambda + \frac{1}{\Gamma} \min_{\beta_1, \beta_2, \omega, \Phi} \Big[ -\omega - \Big( 1 - \alpha^2 + \frac{1}{T_0^2} \sum_{l=1}^{T_0} \sum_{k=1}^{T_0} K(r_l, r_k) \Big) \frac{\beta_1}{2}  \nonumber \\
& \hspace{3cm}  + \Big( 1 + \alpha^2 - \frac{1}{T_0^2} \sum_{l=1}^{T_0} \sum_{k=1}^{T_0} K(r_l, r_k) \Big) \frac{\beta_2}{2} \Big] \Big\} \leq 0 ,\label{eq:min}\\
& \omega\textbf{1} +L^{'} \Phi+\frac{\beta_{1}}{T_{0}} M \textbf{1} +\frac{\beta_{2}}{T_{0}} M \textbf{1}  \leq-
R(w), \label{eq:R1}\\
& \|\Phi\|_{2}^{2}+\beta_{1}^{2} \leq \beta_{2}^{2}. \nonumber
\end{align}
Expression \eqref{eq:min} can be modified as
\begin{equation}
    \min_{\lambda, \beta_1, \beta_2, \omega, \Phi} \Gamma\lambda -\omega - \Big( 1 - \alpha^2 + \frac{1}{T_0^2} \sum_{l=1}^{T_0} \sum_{k=1}^{T_0} K(r_l, r_k) \Big) \frac{\beta_1}{2} + \Big( 1 + \alpha^2 - \frac{1}{T_0^2} \sum_{l=1}^{T_0} \sum_{k=1}^{T_0} K(r_l, r_k) \Big) \frac{\beta_2}{2} \leq 0 . \nonumber
\end{equation}
%Thus, after replacing the constraint \eqref{eq:chance} in the optimization problem CCP-NEW by the above reformulations, we get the following model.
%\begin{align}
%\quad & \min_{c, y, w,\gamma,\lambda,\omega,\beta 1,\beta 2,\Phi} \quad \sum_{k=1}^{m} \theta_k \Big( \gamma_k + \frac{1}{ \delta_k} \sum_{j=1}^{T} c_{jk} p_j \Big) \nonumber\\
%\text{subject to} \quad & -c_{jk} - \gamma_k - \sum_{i=1}^{n} r_{ij} w_i + \|P_j^T w\|_{2}\leq 0, \quad  %k = 1, \ldots, m, \; j = 1, \ldots, T, \nonumber\\
%& c_{jk} \geq 0, \quad  k = 1, \ldots, m, \; j = 1, \ldots, T, \nonumber\\
%& \sum_{i=1}^{n} w_{i} =1, \nonumber\\
%& l_{i} y_{i}\leq w_{i} \leq u_{i} y_{i}, \quad i = 1, %\ldots, n, \nonumber \\
%& y_i \in \{0,1\}, \quad  i = 1, \ldots, n, \nonumber \\
%& \sum_{i=1}^{n} y_i = A, \nonumber\\
%&\nonumber \Gamma\lambda-\omega-\Big(1-\alpha^2+\frac{1}{T_0^{2}} %\sum_{l=1}^{T_{0}}\sum_{k=1}^{T_{0}}K(r_l,r_k)\Big) \frac{\beta_{1}}{2}\\
%& \hspace*{2cm} + \Big(1+\alpha^2-\frac{1}{T_0^{2}} \sum_{l=1}^{T_{0}}\sum_{k=1}^{T_{0}}K(r_l,r_k)\Big) \frac{\beta_{2}}{2} \leq 0,\nonumber\\
%& \omega\textbf{1} +L^{T} \Phi+\frac{\beta_{1}}{T_{0}} M %\textbf{1} +\frac{\beta_{2}}{T_{0}} M \textbf{1}  \leq-
%R(w),\label{eq:R1}\\
%& \|\Phi\|_{2}^{2}+\beta_{1}^{2} \leq %\beta_{2}^{2}.\nonumber
%\end{align}
%Since, $ R(w)=\left(\left[R_{\ast}- \sum_{i=1}^{n} r_{i1} w_i-\lambda\right]^{+}, \ldots,\left[R_{\ast}- \sum_{i=1}^{n} r_{iT} w_i-\lambda\right]^{+}\right)^{'} \label{eq:R(w)}$, 

The constraints \eqref{eq:R1} are equivalent to 
%\begin{equation}
 %   \omega\textbf{1} +L^{T} \Phi+\frac{\beta_{1}}{T_{0}} M \textbf{1} +\frac{\beta_{2}}{T_{0}} M \textbf{1}  \leq-%\left[\left[R_{\ast}- \sum_{i=1}^{n} r_{i1} w_i-%\lambda\right]^{+}, \ldots,\left[R_{\ast}- %\sum_{i=1}^{n} r_{iT} w_i-\lambda\right]^{+}\right]^{T}
%\end{equation}
%By introducing dummy variables $R_j$ $\forall j = %1,\ldots,T$ in the constraint \eqref{eq:R1}
%, We can write it as
\begin{align}
    & \omega\textbf{1} +L^{T} \Phi+\frac{\beta_{1}}{T_{0}} M \textbf{1} +\frac{\beta_{2}}{T_{0}} M \textbf{1}  \leq -\left(R_1 ,\ldots,R_T\right)^{'}, \nonumber\\
    & R_{j} \geq 0 ,\quad j = 1, \ldots , T, \nonumber\\
    & R_{j} \geq R_{\ast}- \sum_{i=1}^{n} r_{ij} w_i-\lambda ,\quad j = 1, \ldots T.\nonumber
\end{align}
Thus, the final problem converts to:

\begin{align}
\text{(RoM-RKHS)} \quad & \min_{c, y, w,\gamma,\lambda,\omega,\beta_1,\beta_2,\Phi} \quad \sum_{k=1}^{m} \theta_k \Big( \gamma_k + \frac{1}{ \delta_k} \sum_{j=1}^{T} c_{jk} p_j \Big) \nonumber\\
\text{subject to} \quad & -c_{jk} - \gamma_k - \sum_{i=1}^{n} r_{ij} w_i + \|P_j^{'} w\|_{2}\leq 0, \quad  k = 1, \ldots, m, \;j = 1, \ldots, T, \nonumber \\
& c_{jk} \geq 0, \quad  k = 1, \ldots, m, \; j = 1, \ldots, T, \nonumber \\
& \sum_{i=1}^{n} w_{i} =1, \nonumber\\
& l_iy_i\leq w_{i} \leq u_{i} y_{i}, \quad i = 1, \ldots, n, \nonumber \\
& y_i \in \{0,1\}, \quad i = 1, \ldots, n, \nonumber \\
& \sum_{i=1}^{n} y_i = A,\label{eq:total assets} \\
&\nonumber \Gamma\lambda-\omega-\Big(1-\alpha^2+\frac{1}{T_0^{2}} \sum_{l=1}^{T_{0}}\sum_{k=1}^{T_{0}}K(r_l,r_k)\Big) \frac{\beta_{1}}{2}\\ 
&\hspace*{2cm}+ \Big(1+\alpha^2-\frac{1}{T_0^{2}} \sum_{l=1}^{T_{0}}\sum_{k=1}^{T_{0}}K(r_l,r_k)\Big) \frac{\beta_{2}}{2} \leq 0, \nonumber\\
& \omega\textbf{1} +L^{'} \Phi+\frac{\beta_{1}}{T_{0}} M \textbf{1} +\frac{\beta_{2}}{T_{0}} M \textbf{1}  \leq-
\left[R_1 ,\ldots,R_T\right]^{'} , \nonumber\\
& R_{j} \geq 0 ,\quad j = 1, \ldots , T, \nonumber\\
& R_{j} \geq R_{\ast}- \sum_{i=1}^{n} r_{ij} w_i-\lambda ,\quad j = 1, \ldots T, \nonumber\\
& \|\Phi\|_{2}^{2}+\beta_{1}^{2} \leq \beta_{2}^{2}, \nonumber\\
&  \lambda,\,\omega,\,\beta_1, \, \beta_2 \in \mathbb{R},\; \Phi \in \mathbb{R}^T.\nonumber 
\end{align}
By taking $A=n$ in constraint \eqref{eq:total assets} we obtain an (RoM-RKHS) model that considers all available assets in the market for investment purposes.

The portfolio on solving (RoM-RKHS) model shall be referred to as RoMP.

\section{Data and Empirical Set-up}\label{section:4}

The experiments are performed on a Windows 10 64-bit operating system with 12 GB RAM and a 12th Gen Intel(R) Core(TM) i7-12700T 1.40 GHz processor. We use MATLAB YALMIP with the MOSEK solver to solve all optimization models.

\subsection{Sample Data and Sample Period}

The sample data for the present empirical analysis consists of weekly closing prices of the constituents for the following six global market indices from September 20, 2012, to September 19, 2024. Data is obtained from the Refinitiv Eikon Datastream. %, resulting in 628 scenarios for each market index:

\begin{enumerate}
    \item \textbf{Data Set 1:} NIKKEI 225 (Japan), 217 assets;
    \item \textbf{Data Set 2:} S\&P 100 (USA), 96 assets;
    \item \textbf{Data Set 3:} NIFTY 50 (India), 48 assets;
    \item \textbf{Data Set 4:} FTSE 100 (UK), 91 assets;
    \item \textbf{Data Set 5:} Dow Jones Industrial Average (USA), 29 assets;
    \item \textbf{Data Set 6:} BOVESPA (Brazil), 63 assets.
\end{enumerate}

The index's stock constituents are periodically revised. Historical price data is only available for stocks that were part of the index as of September 19, 2024. Consequently, data for stocks that failed to survive during the analysis period is absent, leading to survivorship bias in the datasets. The empirical analysis presented serves to demonstrate the superior performance of the proposed models when compared to existing models applied to the same datasets.

Furthermore, alongside the sample data, we evaluate the performance of the proposed models across three distinct market conditions: a general upward (bullish) trend, a neutral trend, and a downward (bearish) trend. To achieve this, we examine the weekly closing values of the Dow Jones Industrial Average (DJIA) and classify the subsequent phases:

\begin{enumerate}
    \item \textbf{Bearish phase:} Oct 02, 2007 - Feb 10, 2009, consisting of 72 weeks.
    \item \textbf{Neutral phase:} Jan 10, 2012 - Jan 01, 2013, consisting of 52 weeks.
    \item \textbf{Bullish phase:} Jan 10, 2017 - Jan 09, 2018, consisting of 53 weeks.
\end{enumerate}
Fig. \eqref{fig:phasebear}-\eqref{fig:phasebull} shows, respectively, bear, neutral, and bull phases of the DJIA in the mentioned periods.

\subsection{Methodology}
A rolling window strategy is applied to test the performance of the proposed robust (RoM-RKHS) model. We establish a 50-week in-sample phase, followed by a 4-week out-of-sample period. We rebalance the portfolio by shifting the in-sample period ahead by four weeks to obtain $T_w= 144$ such windows in the analysis period. Out-of-sample returns are observed for each window and analyzed on 13 performance metrics. The performance comparison of optimal portfolios from the robust model (RoM-RKHS) with $A = 3,6,9,15$ and $ n $, along with benchmark market portfolios and equal weight portfolios, is carried out against their non-robust counterparts. 

The parameters are set to \( m = 3 \), \( \delta_1 = 0.01 \), \( \delta_2 = 0.03 \), and \( \delta_3 = 0.05 \) in MCVaR($R(w)$) to capture extreme losses in the portfolios. The corresponding weights are set as \( \theta_1 = 0.12 \), \( \theta_2 = 0.48 \), and \( \theta_3 = 0.4 \).

All robust models are solved using the in-sample window with \( T = 50\), \(T_0 = 42\) and \( p_j = 1/50, \text{for each } j = 1, \ldots, 50 \), $u_i=0.7$ and $l_i=0.015$, for each $i =1,\ldots,n$. In ellipsoidal support, matrix $P_j$ is $0.072*I_n$ where \( I_n \) is the identity matrix of size \( n\). We set  \( \Gamma = 0.1 \), and \( R_{\ast} \) equal twice the in-sample return when equal weights are allocated to all assets in the portfolio $\Big(R_{\ast} = 2 \sum_{i=1}^{n} \sum_{j=1}^{T} r_{ij}\Big)$. This baseline allocation allows for a reference point to evaluate the relative performance of the optimized portfolio. For better comparison, all parameters, whenever applicable, are taken as the same in the (NoM) model.

\subsection{Performance Measures}

We analyze the out-of-sample performance of all the described models based on the following performance metrics:

1. \textbf{Max Return}: The maximum out-of-sample return observed among all test windows.

2. \textbf{Min Return}: The minimum out-of-sample return observed among all test windows. 

3. \textbf{Mean Return}: The out-of-sample mean return is measured by 
\[
E\left(R_w\right) = \frac{1}{T_w} \sum_{l=1}^{T_w} R_l,
\]where \(R_l\) is the return rate of portfolio in $l-$th window.

4. \textbf{Median Return}: The median of out-of-sample return observed among all test windows.

5. \textbf{Standard Deviation (SD)}: The out-of-sample standard deviation of portfolio returns is computed as 
\[
\sigma_w = \sqrt{\frac{1}{T_w} \sum_{l=1}^{T_w} \left(E(R_w) - R_l\right)^2}.
\]

6. \textbf{Value-at-Risk (VaR) and Conditional Value-at-Risk (CVaR)}: The values for \(VaR_{0.05}\) and \(CVaR_{0.05}\) to measure the downside risks at $5\%$ significance level.

7. \textbf{Stable Tail-Adjusted Return Ratio (STARR)}: STARR is a risk-adjusted reward-risk ratio where risk is measured by \(CVaR_\gamma(R_w)\):

\[
STARR_\gamma = \frac{E(R_w) - r_f}{CVaR_\gamma(R(w))},
\]
where $E(R_w) > r_f$ and $CVaR_\gamma R(w) > 0$, $r_f $ is risk-free return. We select \(\gamma = 0.05\) for the empirical analysis.

8. \textbf{Sharpe Ratio (SR)}: It is defined as the ratio of the excess mean return over the risk-free return \(r_f\) to its standard deviation.
\[
SR = \frac{E(R_w) - r_f}{\sigma_w}, \quad \text{where } E(R_w) > r_f.
\]

9. \textbf{Treynor Ratio}: It measures a portfolio's risk-adjusted return. It is similar to the Sharpe ratio, but it focuses on market risk rather than total risk. It measures how well the portfolio has performed relative to the beta risk:
\[
TREYNOR = \frac{E(R_w) - r_f}{\beta_2},\]

Here, $\beta_2$ is the beta of the portfolio. Which is calculated by regressing out-of-sample returns on market returns.

10. \textbf{Jensen's Alpha}: Jensen's Alpha evaluates the excess return of a portfolio over its expected return, based on its risk and the market's performance:
\[
JENSEN = E(R_w) - [r_f + \beta_2(E(R_M) - r_f)],
\]

where $E(R_M)$ is the expected return of the market.

11. \textbf{Omega Ratio}: The Omega ratio is defined as the ratio of upside deviation from a fixed threshold point \(TP\) (representing reward) to downside deviation (representing risk):

\[
OMEGA = \frac{E\left(\max(TP - R_l, 0)\right)}{E\left(\max(R_l - TP, 0)\right)}.
\]

Here, \(E(TP - R_l)^+\) represents the expected upside deviation above the threshold \(TP\), while \(E(R_l - TP)^+\) represents the expected downside deviation below \(TP\).

12. \textbf{Sortino Ratio}: The Sortino ratio describes the reward-risk ratio associated with the risk below the mean return:
\[
SORTINO = \frac{E(R_w) - r_f}{SSD_w} = \frac{E(R_w) - r_f}{\sqrt{\frac{1}{T_w} \sum_{l=1}^{T_w} \max(E(R_w) - R_l, 0)^2}}, \] 

where  $E(R_w) > r_f$, and \(SSD_w\) is the semi-standard deviation.

For simplicity, we set \(r_f = 0\) in the out-of-sample analysis for all the above metrics across all the datasets.

\section{Empirical Analysis}\label{section:5}
This section compares robust models against nominal models, market indices, and equal weight strategies on their out-of-sample results from six data sets. We also analyze robust model performance across three distinct market trends.

We shall use abbreviated notations for portfolios from the (RoM-RKHS) model, (NoM) model, benchmark market portfolios (benchmark index), and equal-weighted portfolios (naive strategy) by RoMP, NoMP, BMP, and EQP, respectively. 

\subsubsection*{Out-of-sample Analysis on Six Markets Data}

Tables \ref{table:1}–\ref{table:bovespa} present the out-of-sample performance of the portfolios on the performance metrics outlined in Section 6 for the four values of $A$ = 3, 6, 9, and 15, as well as when $A=n$.

In Table \ref{table:1}, the NIKKEI 225, RoMP yield higher mean returns than  NoMP, BMP, EQP. The risk metrics (SD, VaR, CVaR) are also lower for RoMP than the others. The reward-to-risk ratios, including the Sortino, Omega, Treynor, and STARR ratios, are all superior for RoMP with $A$ = 6. Moreover, the robust model (RoM-RKHS) shows a more favourable Jensen’s alpha for $A$ = 6. The Sharpe ratio is slightly higher for $A=n$ and is comparable to $A$ = 6 in a robust model.

In Table \ref{table:sp}, the S\&P 100, RoMP has higher mean returns than NoMP; however, other than $A$ = 9, RoMP outperforms the EQP. Compared to other portfolios, the SD, VaR, and CVaR are also low in RoMP, respectively, for $A=n$, $A$ = 6 and 15. The reward-to-risk ratios outperform for RoMP with $A$ = 9. Moreover, the robust model shows a more favourable Jensen’s alpha for $A$ = 9. At the same time, the omega ratio is higher in BMP.

In Table \ref{table:nifty}, the NIFTY 50, RoMP outshines all other portfolios with higher mean returns. In addition, the VaR and CVaR are low in RoMP for $A$ = 9 compared to others. The SD is the lowest in BMP. The reward-to-risk ratios and alpha value are superior for RoMP with $A$ = 9.  The omega ratio is higher for RoMP with $A=n$.

In Table \ref{table:ftse}, the FTSE 100, the RoMP generally yield higher mean returns than NoMP, BMP, EQP. The RoMP with $A$ = 6 provides higher mean returns than other models. The robust model with $A$ = 9 generates low risk. The BMP has the lowest SD. However, the reward-to-risk ratios and an alpha value ($A$ = 6) in RoMP are superior, while the omega ratio is highest for NoMP with $A=n$.

In Table \ref{table:dow}, the DJIA 30, the RoMP generally exhibit higher mean returns than the others. The RoMP with $A$ = 9 has the highest mean return, outperforming NoMP and BMP. The RoMP with $A$ = 9 provides the highest alpha, lower risk, and superior reward-to-risk ratios. The NoMP and BMP generally display higher risks and lower performance metrics across key indicators.

Table \ref{table:bovespa}, the BOVESPA market, the RoMP surpass in mean returns. The RoMP with $A$ = 15 achieves the highest mean return, which yields a higher return than the maximum return from any NoMP. The NoMP with $A$ = 3 produces the lowest mean return, while RoMP with $A$ = 3 provides a substantially higher return. The RoMP with $A$ = 9 has the lowest SD, showing less volatility than the other portfolios. The NoMP with $A$ = 15 has the lowest VaR (0.05), while the RoMP with $A$ = 9 offers a higher but competitive VaR (0.05). The CVaR is lower for the RoMP with $A$ = 9. 

Jensen's alpha is highest for RoMP with $A$ = 15, showing that the (RoM-RKHS) model has the best performance relative to its expected return.  Jensen's alpha is negative for NoMP with $A$ = 3, indicating underperformance relative to the market benchmark. The RoMP with $K$ = 9 provides the highest Sharpe, Treynor, STARR, and Omega ratios, while it with $K$ = 15 delivers the highest Sortino ratio.

{\footnotesize{
\begin{table}[!ht]
   \setlength{\tabcolsep}{1.5pt}
    \centering
    \scalebox{0.8}{
    \begin{tabular}{|c|c|c|c|c|c|c|c|c|c|c|c|c|}
    \hline
    & \multicolumn{5}{c|}{\textbf{(NoM)}} && \multicolumn{5}{c|}{\textbf{(RoM-RKHS)}}& \\
    \hline
    \textbf{Statistics} & \textbf{3} & \textbf{6} & \textbf{9} & \textbf{15} & \textbf{n} & \textbf{BMP} & \textbf{3} & \textbf{6} & \textbf{9} & \textbf{15} & \textbf{n} & \textbf{EQP} \\
    & \textbf{Assets} & \textbf{Assets} & \textbf{Assets} & \textbf{Assets} & \textbf{Assets} &  & \textbf{Assets} & \textbf{Assets} & \textbf{Assets} & \textbf{Assets} & \textbf{Assets} &  \\
 \hline
\textbf{MAX} & 33.06 & 38.154 & 40.891 & 38.869 & 40.396 & 36.5 & 41.875 & \textbf{45.131} & 29.051 & 29.051 & 26.941 & 37.473 \\
\hline
\textbf{MIN} & -65.132 & -43.191 & -63.488 & -62.287 & -63.45 & \textbf{-59.5} & -70.004 & -66.144 & -60.264 & -60.264 & -60.257 & -67.444 \\
\hline
\textbf{MEAN} & 1.976 & 2.727 & 2.601 & 2.579 & 2.529 & 2.054 & 2.417 & \textbf{3.079} & 2.614 & 2.602 & 2.67 & 2.246 \\
\hline
\textbf{MEDIAN} & 2.555 & 4.201 & 3.711 & 4.137 & 4.183 & 3.15 & 2.726 & 4.376 & 3.962 & 3.962 & \textbf{4.447} & 4.041 \\
\hline
\textbf{SD} & 13.751 & 12.894 & 12.995 & 12.618 & 12.899 & 12.365 & 15.263 & 13.674 & 12.061 & 12.064 & \textbf{11.832} & 12.828 \\
\hline
\textbf{VAR 0.05} & 21.526 & 17.263 & 17.851 & 17.744 & 18.097 & 18.78 & 24.041 & 17.533 & 15.176 & 15.176 & \textbf{14.868} & 18.168 \\
\hline
\textbf{CVAR 0.05} & 32.891 & 31.381 & 32.956 & 32.343 & 32.707 & 31.629 & 37.389 & 32.556 & 30.717 & 30.744 & \textbf{30.344} & 33.088 \\
\hline
\textbf{STARR 0.05} & 60.076 & 86.89 & 78.915 & 79.726 & 77.309 & 64.947 & 64.636 & \textbf{94.57} & 85.104 & 84.644 & 87.981 & 67.874 \\
\hline
\textbf{SHARPE} & 143.7 & 211.5 & 200.15 & 204.39 & 196.06 & 166.11 & 158.36 &	225.17	 & 216.73 & 215.68 & \textbf{229.88}& 175.09\\
\hline
\textbf{TREYNOR} & 2.646 & 3.557 & 3.201 & 3.145 & 3.069 & 2.054 & 2.97 & \textbf{3.929} & 3.224 & 3.211 & 3.332 & 2.285 \\
\hline
\textbf{JENSEN} & 0.442 & 1.152 & 0.932 & 0.894 & 0.836 & 0 & 0.745 & \textbf{1.469} & 0.949 & 0.938 & 1.024 & 0.227 \\
\hline
\textbf{OMEGA} & 1020 & 1013.4 & 975.07 & 953.42 & 906.05 & 999.61 & 848.7 & \textbf{1057.8} & 784.11 & 782.34 & 801.25 & 943.13 \\
\hline
\textbf{SORTINO} & 130.96 & 192.24 & 171.94 & 172.58 & 164.4 & 146.92 & 136.39 & \textbf{199.27} & 170.04 & 169.13 & 175.55 & 146.36 \\
\hline
    \end{tabular}}
    \vspace*{0.3cm}
    \caption{The out-of-sample statistics (* $10^{-3}$) for NIKKEI 225 dataset.}\label{table:1}
\end{table}
}}

{\footnotesize{
\begin{table}[!ht]
   \setlength{\tabcolsep}{1.5pt}
    \centering
    \scalebox{0.8}{
    \begin{tabular}{|c|c|c|c|c|c|c|c|c|c|c|c|c|}
    \hline
    & \multicolumn{5}{c|}{\textbf{(NoM)}} && \multicolumn{5}{c|}{\textbf{(RoM-RKHS)}}& \\
    \hline
    \textbf{Statistics} & \textbf{3} & \textbf{6} & \textbf{9} & \textbf{15} & \textbf{n} & \textbf{BMP} & \textbf{3} & \textbf{6} & \textbf{9} & \textbf{15} & \textbf{n} & \textbf{EQP} \\
    & \textbf{Assets} & \textbf{Assets} & \textbf{Assets} & \textbf{Assets} & \textbf{Assets} &  & \textbf{Assets} & \textbf{Assets} & \textbf{Assets} & \textbf{Assets} & \textbf{Assets} &  \\
 \hline
\textbf{MAX} & 28.876 & 28.636 & 28.492 & 28.442 & 28.554 & 32.7 & 29.833 & 32.132 & 32.354 & 30.108 & 30.246 & \textbf{38.097} \\
\hline
\textbf{MIN} & -59.329 & -60.92 & -54.62 & -57.14 & -55.585 & -68.3 & \textbf{-53.885} & -61.615 & -57.25 & -59.474 & -58.179 & -73.663 \\
\hline
\textbf{MEAN} & 1.277 & 1.102 & 1.705 & 1.579 & 1.552 & 2.519 & 1.749 & 2.104 & \textbf{3.014} & 2.629 & 2.732 & 2.812 \\
\hline
\textbf{MEDIAN} & 1.795 & 2.028 & 2.279 & 2.783 & 2.324 & 4.4 & 2.171 & 3.519 & \textbf{4.718} & 4.16 & 4.561 & 4.135 \\
\hline
\textbf{SD} & 10.425 & 10.291 & 9.962 & 9.891 & 10.035 & 10.93 & 11.408 & 10.621 & 10.223 & 9.959 & \textbf{9.886} & 11.175 \\
\hline
\textbf{VAR 0.05} & 16.347 & 15.02 & 15.077 & 15.116 & 15.05 & 16.51 & 17.786 & \textbf{13.079} & 13.244 & 14.179 & 13.589 & 15.32 \\
\hline
\textbf{CVAR 0.05} & 25.704 & 26.564 & 24.14 & 24.805 & 24.572 & 28.643 & 27.112 & 24.298 & 23.151 & \textbf{23.143} & 23.888 & 28.393 \\
\hline
\textbf{STARR 0.05} & 49.687 & 41.483 & 70.612 & 63.676 & 63.147 & 87.936 & 64.493 & 86.584 & \textbf{130.2} & 113.62 & 114.35 & 99.056 \\
\hline
\textbf{SHARPE} & 122.49&107.08&	171.15&	159.64&	154.66 & 230.47 & 153.31&198.1	&\textbf{294.83}	&263.98
 & 276.35
 & 251.63 \\
\hline
\textbf{TREYNOR} & 1.936 & 1.55 & 2.407 & 2.2 & 2.173 & 2.519 & 2.317 & 2.842 & \textbf{3.963} & 3.423 & 3.507 & 2.895 \\
\hline
\textbf{JENSEN} & -0.385 & -0.689 & -0.079 & -0.229 & -0.247 & 0 & -0.152 & 0.239 & \textbf{1.099} & 0.694 & 0.77 & 0.366 \\
\hline
\textbf{OMEGA} & 1025.6 & 841.84 & 802.91 & 855.92 & 947.56 & \textbf{1093.3} & 884.17 & 1000.1 & 1004.7 & 957.3 & 936.1 & 898.34 \\
\hline
\textbf{SORTINO} & 109.17 & 88.025 & 138.54 & 129.81 & 132.93 & 190.07 & 130.86 & 171.98 & \textbf{244.9} & 213 & 217.43 & 189.19 \\
 \hline
    \end{tabular}}
    \vspace*{0.3cm}
    \caption{The out-of-sample statistics (* $10^{-3}$) for S\&P 100 dataset.}\label{table:sp}
    
\end{table}

{\footnotesize{
\begin{table}[!ht]
   \setlength{\tabcolsep}{1.5pt}
    \centering
    \scalebox{0.8}{
    \begin{tabular}{|c|c|c|c|c|c|c|c|c|c|c|c|c|}
    \hline
    & \multicolumn{5}{c|}{\textbf{(NoM)}} && \multicolumn{5}{c|}{\textbf{(RoM-RKHS)}}& \\
    \hline
    \textbf{Statistics} & \textbf{3} & \textbf{6} & \textbf{9} & \textbf{15} & \textbf{n} & \textbf{BMP} & \textbf{3} & \textbf{6} & \textbf{9} & \textbf{15} & \textbf{n} & \textbf{EQP} \\
    & \textbf{Assets} & \textbf{Assets} & \textbf{Assets} & \textbf{Assets} & \textbf{Assets} &  & \textbf{Assets} & \textbf{Assets} & \textbf{Assets} & \textbf{Assets} & \textbf{Assets} &  \\
 \hline
\textbf{MAX} & 38.907 & 37.664 & 37.029 & 35.306 & 38.818 & 25.6 & \textbf{44.891} & 40.12 & 37.402 & 38.541 & 37.538 & 27.232 \\
\hline
\textbf{MIN} & \textbf{-36.032} & -41.682 & -41.653 & -40.02 & -42 & -55.9 & -39.885 & -41.006 & -44.532 & -43.078 & -42.47 & -58.221 \\
\hline
\textbf{MEAN} & 3.679 & 2.985 & 3.287 & 3.143 & 3.277 & 3.262 & \textbf{5.158} & 5.014 & 5.115 & 5.006 & 4.884 & 4.407 \\
\hline
\textbf{MEDIAN} & 4.228 & 3.951 & 4.149 & 4.041 & 4.149 & 4.3 & \textbf{5.839} & 5.592 & 5.46 & 5.085 & 5.23 & 4.717 \\
\hline
\textbf{SD} & 13.268 & 12.687 & 12.223 & 11.914 & 12.281 & \textbf{11.12} & 14.148 & 12.678 & 11.588 & 11.453 & 11.759 & 11.626 \\
\hline
\textbf{VAR 0.05} & 18.293 & 19.368 & 17.425 & 17.037 & 17.803 & 17.59 & 18.865 & 15.888 & \textbf{13.777} & 15.315 & 16.382 & 16.157 \\
\hline
\textbf{CVAR 0.05} & 26.86 & 25.897 & 25.342 & 25.105 & 25.163 & 26.229 & 28.917 & 25.747 & \textbf{23.742} & 23.824 & 25.097 & 26.176 \\
\hline
\textbf{STARR 0.05} & 136.96 & 115.27 & 129.69 & 125.21 & 130.25 & 124.36 & 178.39 & 194.75 & \textbf{215.44} & 210.12 & 194.59 & 168.37 \\
\hline
\textbf{SHARPE} & 277.28
 &235.28&	268.92&	263.81&	266.83 & 293.35
 &364.57	&395.49&	\textbf{441.4}&	437.09
 & 415.34
 & 379.06 \\
\hline
\textbf{TREYNOR} & 4.589 & 3.522 & 3.93 & 3.751 & 3.918 & 3.262 & 6.01 & 5.972 & \textbf{6.25} & 5.959 & 5.648 & 4.294 \\
\hline
\textbf{JENSEN} & 1.064 & 0.221 & 0.559 & 0.41 & 0.549 & 0 & 2.359 & 2.275 & \textbf{2.446} & 2.266 & 2.063 & 1.06 \\
\hline
\textbf{OMEGA} & 1218.8 & 1161.2 & 1170.4 & 1168 & 1216.5 & 1040.2 & 1108.7 & 1031.5 & 1260.9 & 1176.7 & \textbf{1285.1} & 1084 \\
\hline
\textbf{SORTINO} & 283.9 & 233.53 & 266.41 & 258.41 & 269.13 & 252.02 & 354.88 & 369.69 & \textbf{420.69} & 406.32 & 401.45 & 325.96 \\
\hline 
    \end{tabular}}
    \vspace*{0.3cm}
    \caption{The out-of-sample statistics (* $10^{-3}$) for NIFTY 50 dataset.}\label{table:nifty}
    
\end{table}
}}
{\footnotesize{
\begin{table}[!ht]
   \setlength{\tabcolsep}{1.5pt}
    \centering
    \scalebox{0.8}{
    \begin{tabular}{|c|c|c|c|c|c|c|c|c|c|c|c|c|}
    \hline
    & \multicolumn{5}{c|}{\textbf{(NoM)}} && \multicolumn{5}{c|}{\textbf{(RoM-RKHS)}}& \\
    \hline
    \textbf{Statistics} & \textbf{3} & \textbf{6} & \textbf{9} & \textbf{15} & \textbf{n} & \textbf{BMP} & \textbf{3} & \textbf{6} & \textbf{9} & \textbf{15} & \textbf{n} & \textbf{EQP} \\
    & \textbf{Assets} & \textbf{Assets} & \textbf{Assets} & \textbf{Assets} & \textbf{Assets} &  & \textbf{Assets} & \textbf{Assets} & \textbf{Assets} & \textbf{Assets} & \textbf{Assets} &  \\
 \hline
\textbf{MAX} & 27.739 & 26.882 & 26.397 & 24.985 & 27.739 & 30.2 & 22.669 & 24.835 & \textbf{35.573} & 30.449 & 30.449 & 30.534 \\
\hline
\textbf{MIN} & \textbf{-65.218} & -70.246 & -72.811 & -73.509 & -72.811 & -80.1 & -58.121 & -76.53 & -74.316 & -72.365 & -72.332 & -80.234 \\
\hline
\textbf{MEAN} & 1.594 & 1.842 & 1.531 & 1.485 & 1.577 & 0.663 & 1.622 & \textbf{1.862} & 1.545 & 1.832 & 1.774 & 1.75 \\
\hline
\textbf{MEDIAN} & 2.241 & 2.501 & 2.051 & 2.23 & 2.499 & 1.3 & 2.412 & 2.275 & 1.711 & \textbf{2.7} & 2.341 & 2.341 \\
\hline
\textbf{SD} & 10.489 & 10.62 & 10.264 & 10.304 & 10.152 & 10.494 & 10.143 & 10.596 & 9.902 & \textbf{9.785} & 9.803 & 11.541 \\
\hline
\textbf{VAR 0.05} & 15.215 & 14.626 & 13.343 & 13.173 & 12.571 & 13.83 & 16.123 & 12.011 & \textbf{11.278} & 12.284 & 12.562 & 14.273 \\
\hline
\textbf{CVAR 0.05} & 26.895 & 26.591 & 26.039 & 26.1 & 25.479 & 27.057 & 26.21 & 24.133 & \textbf{22.452} & 22.591 & 22.695 & 27.494 \\
\hline
\textbf{STARR 0.05} & 59.265 & 69.269 & 58.782 & 56.91 & 61.903 & 24.511 & 61.881 & 77.174 & 68.835 & \textbf{81.112} & 78.172 & 63.631 \\
\hline
\textbf{SHARPE} & 151.97
 & 173.45&	149.16	&144.12&155.34& 63.18
&159.91	&175.73&	156.03&	\textbf{187.23} & 180.96& 151.63
 \\
\hline
\textbf{TREYNOR} & 2.258 & 2.597 & 2.08 & 1.926 & 2.157 & 0.663 & \textbf{2.762} & 2.46 & 2.043 & 2.285 & 2.211 & 1.726 \\
\hline
\textbf{JENSEN} & 1.126 & 1.371 & 1.043 & 0.974 & 1.092 & 0 & 1.232 & \textbf{1.36} & 1.044 & 1.301 & 1.242 & 1.077 \\
\hline
\textbf{OMEGA} & 1002.3 & 917.75 & 882.88 & 924.19 & \textbf{1037.8} & 873.03 & 785.73 & 980.17 & 953.72 & 1017.1 & 910.64 & 916.19 \\
\hline
\textbf{SORTINO} & 128.45 & 140.1 & 117.34 & 115.35 & 128.62 & 50.968 & 125.55 & 144.09 & 125.73 & \textbf{151.27} & 141.5 & 123.88 \\
\hline
    \end{tabular}}
    \vspace*{0.3cm}
    \caption{The out-of-sample statistics (* $10^{-3}$) for FTSE 100 dataset.}\label{table:ftse}
    
\end{table}
}}
{\footnotesize{
\begin{table}[!ht]
   \setlength{\tabcolsep}{1.5pt}
    \centering
    \scalebox{0.8}{
    \begin{tabular}{|c|c|c|c|c|c|c|c|c|c|c|c|c|}
    \hline
    & \multicolumn{5}{c|}{\textbf{(NoM)}} && \multicolumn{5}{c|}{\textbf{(RoM-RKHS)}}& \\
    \hline
    \textbf{Statistics} & \textbf{3} & \textbf{6} & \textbf{9} & \textbf{15} & \textbf{n} & \textbf{BMP} & \textbf{3} & \textbf{6} & \textbf{9} & \textbf{15} & \textbf{n} & \textbf{EQP} \\
    & \textbf{Assets} & \textbf{Assets} & \textbf{Assets} & \textbf{Assets} & \textbf{Assets} &  & \textbf{Assets} & \textbf{Assets} & \textbf{Assets} & \textbf{Assets} & \textbf{Assets} &  \\
 \hline
\textbf{MAX} & 31.788 & 31.052 & 30.865 & 29.807 & 32.002 & 31.8 & \textbf{33.457} & 30.633 & 29.379 & 30.292 & 29.306 & 39.882 \\
\hline
\textbf{MIN} & -51.325 & -56.925 & -54.356 & -54.579 & -54.748 & -74.9 & -57.062 & -54.009 & \textbf{-51.655} & -57.408 & -56.542 & -68.897 \\
\hline
\textbf{MEAN} & 2.497 & 1.712 & 1.756 & 1.763 & 1.769 & 2.046 & 2.523 & 2.723 & \textbf{2.838} & 2.201 & 2.545 & 2.538 \\
\hline
\textbf{MEDIAN} & 3.805 & 2.713 & 2.377 & 2.959 & 2.15 & 2.5 & 3.142 & 2.964 & 3.202 & 2.161 & 3.534 & \textbf{3.659} \\
\hline
\textbf{SD} & 10.856 & 10.699 & 10.465 & 10.321 & 10.648 & 10.762 & 10.492 & 10.062 & \textbf{9.896} & 10.077 & 11.071 & 10.54 \\
\hline
\textbf{VAR 0.05} & 13.895 & 14.288 & 14.038 & 13.45 & 14.693 & 14.8 & \textbf{11.237} & 13.147 & 12.884 & 14.017 & 14.472 & 13.692 \\
\hline
\textbf{CVAR 0.05} & 21.988 & 22.728 & 22.522 & 22.197 & 22.773 & 29.017 & 22.209 & \textbf{20.661} & 21.199 & 22.656 & 27.383 & 25.909 \\
\hline
\textbf{STARR 0.05} & 113.54 & 75.324 & 77.979 & 79.41 & 77.662 & 70.505 & 113.62 & 131.8 & \textbf{133.89} & 97.138 & 92.929 & 97.967 \\
\hline
\textbf{SHARPE} & 230.01
 & 160.01&167.8&170.82	&166.13& 190.11& 240.47	&270.62	&\textbf{286.78}	&218.42
 & 229.88
 & 240.8
 \\
\hline
\textbf{TREYNOR} & 3.543 & 2.239 & 2.333 & 2.31 & 2.33 & 2.046 & 3.627 & 3.622 & \textbf{3.726} & 2.744 & 3.172 & 2.628 \\
\hline
\textbf{JENSEN} & 1.055 & 0.147 & 0.216 & 0.201 & 0.215 & 0 & 1.1 & 1.185 & \textbf{1.28} & 0.56 & 0.903 & 0.562 \\
\hline
\textbf{OMEGA} & 1110.5 & 981.31 & 1000.7 & 922.43 & 1025.9 & 986.2 & \textbf{1181.7} & 1060.6 & 1101.8 & 1030.4 & 844.44 & 1148.8 \\
\hline
\textbf{SORTINO} & 222.48 & 147.51 & 155.47 & 152.86 & 156.16 & 152.66 & 228.61 & 245.64 & \textbf{258.76} & 178.49 & 177.45 & 202.06 \\
\hline
    \end{tabular}}
    \vspace*{0.3cm}
    \caption{The out-of-sample statistics (* $10^{-3}$) for DJIA dataset.}\label{table:dow}
    
\end{table}
}}

{\footnotesize{
\begin{table}[!ht]
   \setlength{\tabcolsep}{1.5pt}
    \centering
    \scalebox{0.8}{
    \begin{tabular}{|c|c|c|c|c|c|c|c|c|c|c|c|c|}
    \hline
    & \multicolumn{5}{c|}{\textbf{(NoM)}} && \multicolumn{5}{c|}{\textbf{(RoM-RKHS)}}& \\
    \hline
    \textbf{Statistics} & \textbf{3} & \textbf{6} & \textbf{9} & \textbf{15} & \textbf{n} & \textbf{BMP} & \textbf{3} & \textbf{6} & \textbf{9} & \textbf{15} & \textbf{n} & \textbf{EQP} \\
    & \textbf{Assets} & \textbf{Assets} & \textbf{Assets} & \textbf{Assets} & \textbf{Assets} &  & \textbf{Assets} & \textbf{Assets} & \textbf{Assets} & \textbf{Assets} & \textbf{Assets} &  \\
 \hline
\textbf{MAX} & 31.52 & 121.03 & \textbf{128.37} & 101.47 & 125.04 & 48.6 & 39.196 & 40.323 & 39.63 & 38.475 & 39.395 & 53.168 \\
\hline
\textbf{MIN} & \textbf{-75.751} & -88.162 & -98.322 & -91.81 & -91.81 & -101.7 & -89.797 & -89.445 & -90.404 & -90.443 & -89.08 & -94.502 \\
\hline
\textbf{MEAN} & 0.323 & 1.649 & 1.119 & 1.002 & 1.138 & 1.305 & 2.155 & 2.192 & 2.232 & 2.248 & 2.247 & \textbf{2.624} \\
\hline
\textbf{MEDIAN} & 0.407 & 0.423 & 0.984 & 1.153 & 0 & 1.35 & \textbf{3.428} & 3.195 & 3.095 & 3.16 & 3.276 & 3.33 \\
\hline
\textbf{SD} & 14.096 & 18.914 & 18.014 & 16.106 & 17.49 & 16.707 & 13.908 & 13.799 & \textbf{13.77} & 13.885 & 13.882 & 16.751 \\
\hline
\textbf{VAR 0.05} & 18.898 & 21.06 & 18.436 & \textbf{17.418} & 17.981 & 18.9 & 19.005 & 19.208 & 18.958 & 19.261 & 19.149 & 19.633 \\
\hline
\textbf{CVAR 0.05} & 41.229 & 40.401 & 42.327 & 39.547 & 40.323 & 45.383 & 33.625 & 33.185 & \textbf{33.118} & 33.391 & 33.441 & 39.46 \\
\hline
\textbf{STARR 0.05} & 7.827 & 40.813 & 26.429 & 25.341 & 28.216 & 28.752 & 64.078 & 66.048 & \textbf{67.407} & 67.321 & 67.206 & 66.495 \\
\hline
\textbf{SHARPE} & 22.91
 & 87.18&	62.12&	62.21&	65.07 & 78.11 & 154.95&158.85	&\textbf{162.09}&	161.9 & 161.86
 & 156.65
 \\
\hline
\textbf{TREYNOR} & 0.699 & 3.003 & 1.926 & 1.742 & 2.058 & 1.305 & 3.142 & 3.21 & \textbf{3.279} & 3.263 & 3.271 & 2.718 \\
\hline
\textbf{JENSEN} & -0.28 & 0.932 & 0.361 & 0.251 & 0.416 & 0 & 1.26 & 1.301 & 1.344 & 1.349 & 1.351 & \textbf{1.364} \\
\hline
\textbf{OMEGA} & 790.08 & 1138.2 & 1041.4 & 865.15 & 1312.7 & 998.57 & 1041.8 & 1026.7 & 1007 & 1062.4 & 1033.3 & \textbf{1137} \\
\hline

\textbf{SORTINO} & 18.9 & 100.11 & 64.147 & 58.327 & 65.784 & 70.364 & 133.57 & 136.09 & 136.93 & 139.76 & 139.25 & \textbf{148.37} \\
\hline
    \end{tabular}}
    \vspace*{0.3cm}
    \caption{The out-of-sample statistics (* $10^{-3}$) for BOVESPA dataset.}\label{table:bovespa}
    
\end{table}
}}
\clearpage
For illustration, Fig. \ref{fig:nikkei225} - \ref{fig:BOVESPA} present the weekly cumulative out-of-sample returns for portfolios employing the (NoM), (RoM-RKH) models, alongside BMP and EQP, across each dataset. The out-of-sample returns from each rolling window are consolidated into a single out-of-sample return series, creating these plots. The cumulative out-of-sample returns are subsequently calculated across 144 distinct windows. The data provides returns comparisons of portfolios from various models.

From Fig. \ref{fig:nikkei225}, we note that RoMP with $A$ = 6 gives maximum rewards compared to NoMP in the Nikkei market. And NoMP with $A$ = 6 is more rewarding than BMP and EQP.

From Fig. \ref{fig:sp100}, the RoMP with $A$ = 9 gives the maximum reward in the S\&P market. However, the NoMP with $A$ = 9 is suboptimal compared to the BMP and EQP. EQP gives a nearly equal return as RoMP with $A$ = 9.

From Fig. \ref{fig:nifty50}, in the NIFTY 50 market, RoMP with $A$ = 3 gives maximum rewards; NoMP with $A$ = 9 is suboptimal to EQP. However, it gives returns nearly equal to the BMP.

From Fig. \ref{fig:ftse100},  RoMP with $A$ = 6 gives maximum rewards in the FTSE 100 market.

From Fig. \ref{fig:DOW}, RoMP with $A$ = 9 gives maximum rewards in the DJIA market.
However, the EQP outperformed NoMP and BMP.

The EQP outperformed other portfolios from Fig. \ref{fig:BOVESPA}. However, the RoMP with $A$ = 15 gives better cumulative returns than NoMP and BMP.

\begin{figure}[htbp]
    \centering
    \includegraphics[width=1\textwidth]{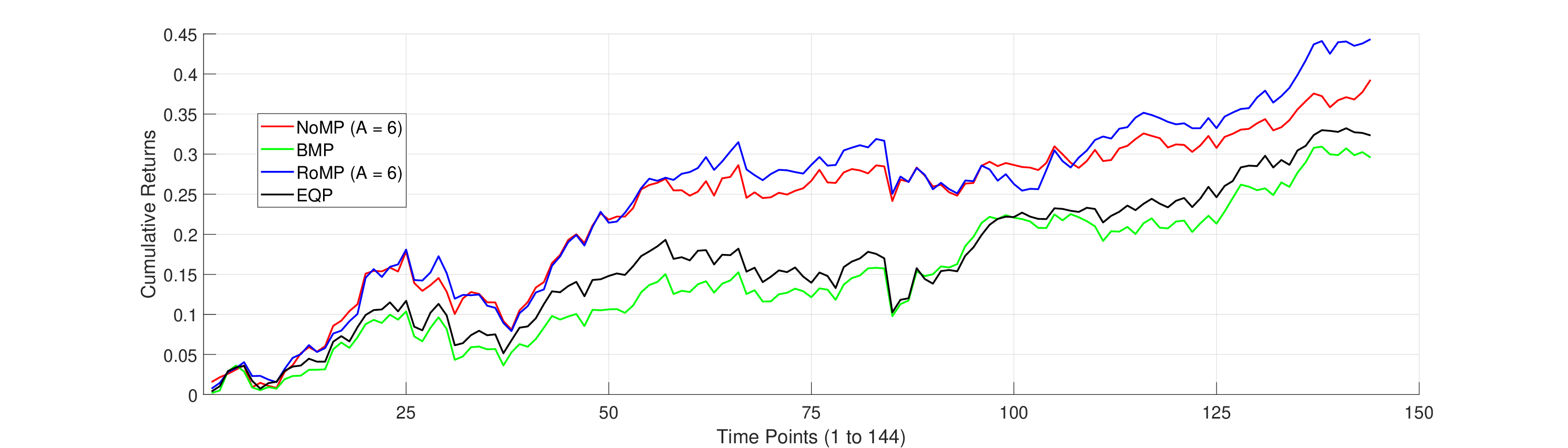}
    \caption{Cumulative returns from portfolios on Nikkei 225}
    \label{fig:nikkei225}
\end{figure}

\begin{figure}[htbp]
    \centering
    \includegraphics[width=1\textwidth]{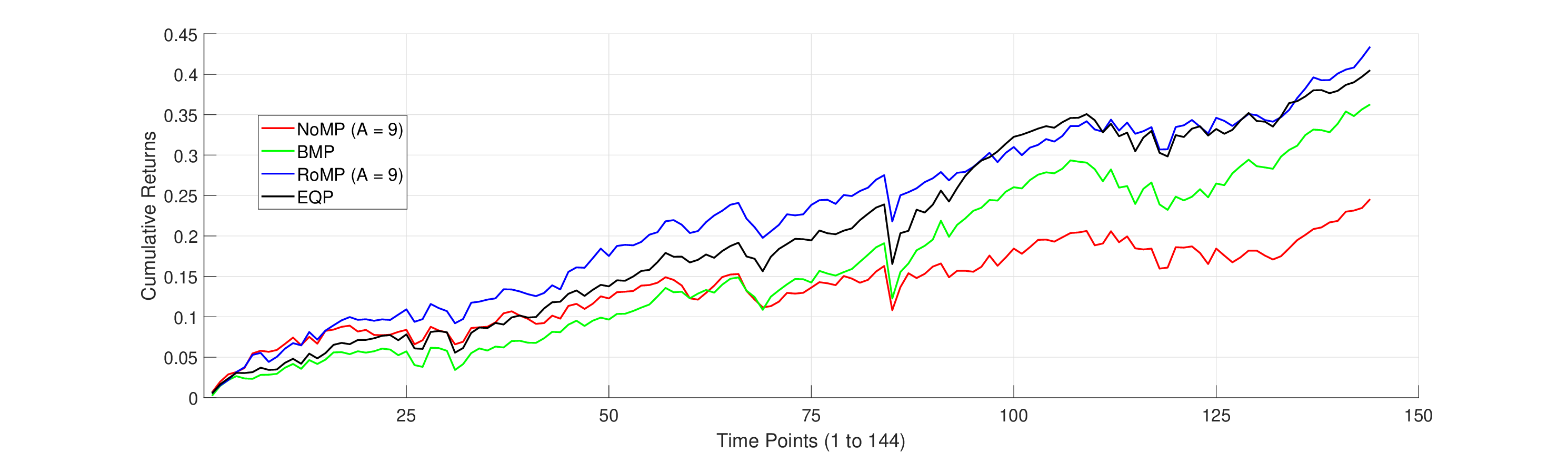}
    \caption{Cumulative returns from portfolios on S\&P 100}
    \label{fig:sp100}
\end{figure}

\begin{figure}[htbp]
    \centering
    \includegraphics[width=1\textwidth]{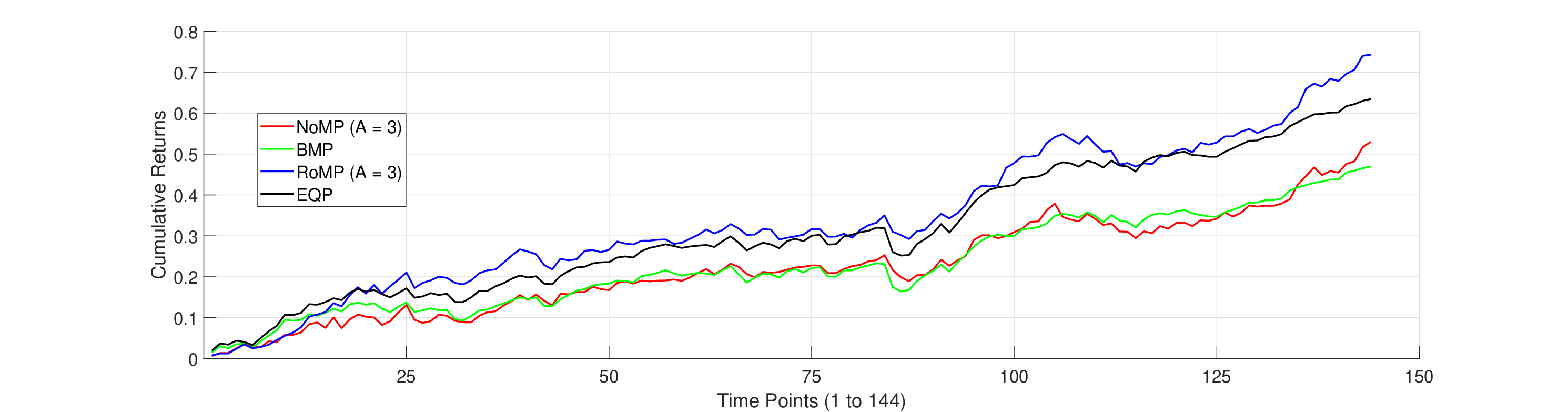}
    \caption{Cumulative returns from portfolios on Nifty 50}
    \label{fig:nifty50}
\end{figure}

\begin{figure}[htbp]
    \centering
    \includegraphics[width=1\textwidth]{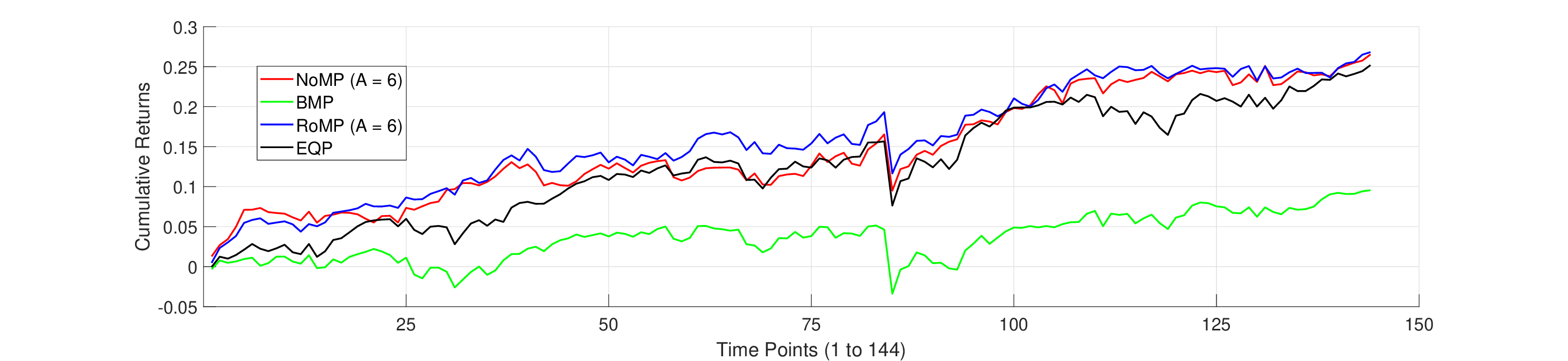}
    \caption{Cumulative returns from portfolios on FTSE 100}
    \label{fig:ftse100}
\end{figure}

\begin{figure}[htbp]
    \centering
    \includegraphics[width=1\textwidth]{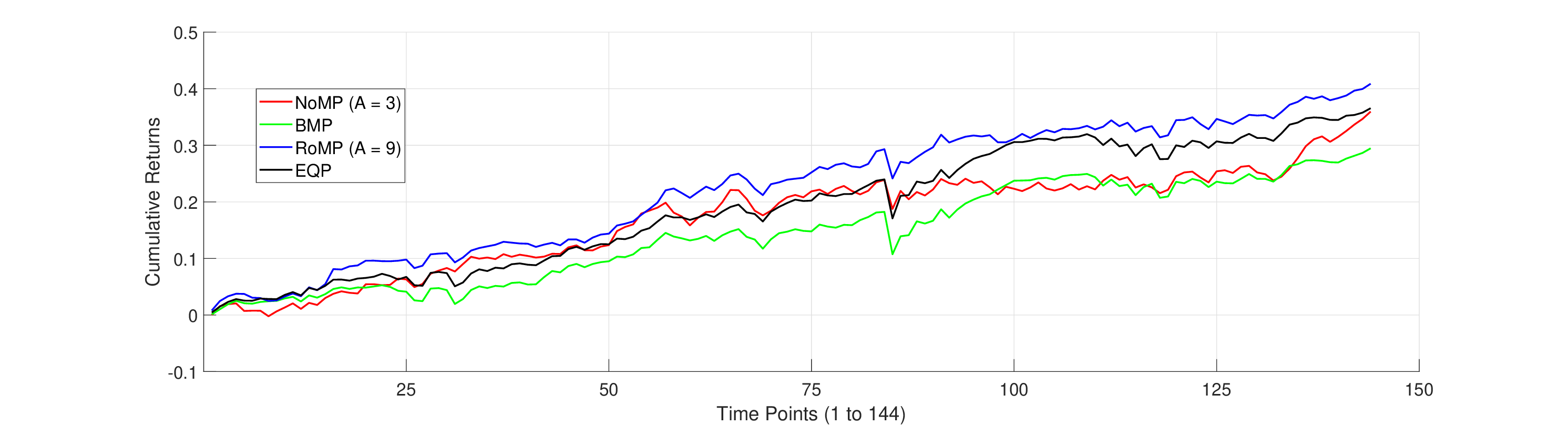} 
    \caption{Cumulative returns from portfolios on DJIA}
    \label{fig:DOW}
\end{figure}

\begin{figure}[htbp]
    \centering
    \includegraphics[width=1\textwidth]{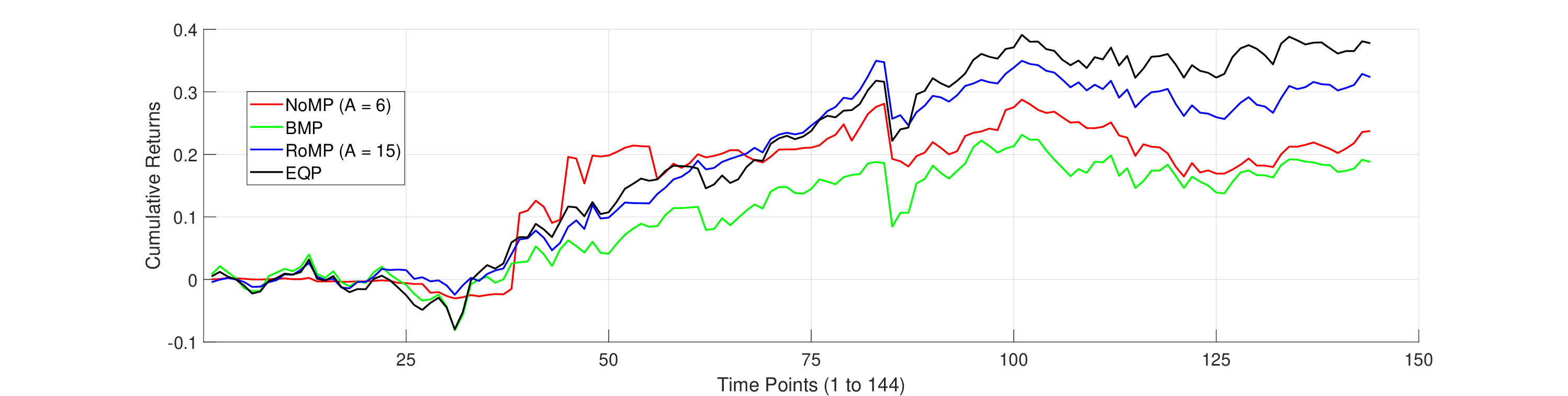} 
    \caption{Cumulative returns from portfolios on BOVESPA.}
    \label{fig:BOVESPA}
\end{figure}
\clearpage

\subsubsection*{Out-of-sample Analysis in Different Market Phases}
For analyzing the performance of the proposed model, especially in the bearish phases of the market, we consider DJIA weekly data as stated in subsection 4.1.

Fig. \ref{fig:phasebear}-\ref{fig:phasebull} demonstrate the three different phases of DJIA under consideration.

\begin{figure}[!htbp]
    \centering
    \includegraphics[width=1\textwidth]{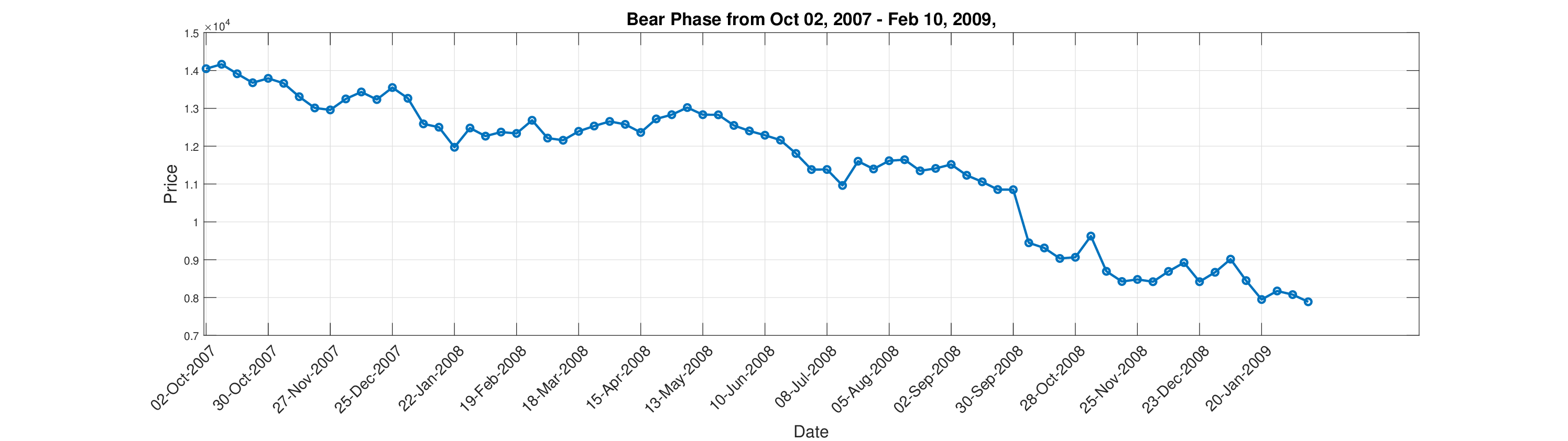}
    \caption{Bearish trend.}
    \label{fig:phasebear}
\end{figure}
\begin{figure}[!htbp]
    \centering
    \includegraphics[width=1\textwidth]{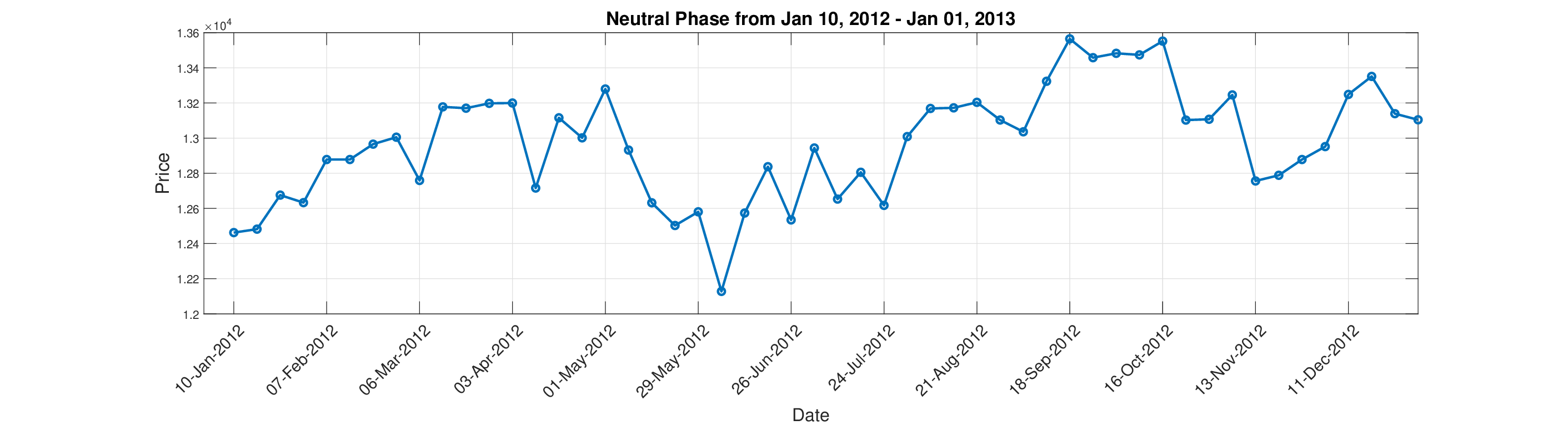}
    \caption{Neutral trend.}
    \label{fig:phaseneutral}
\end{figure}
\begin{figure}[!htbp]
    \centering
    \includegraphics[width=1\textwidth]{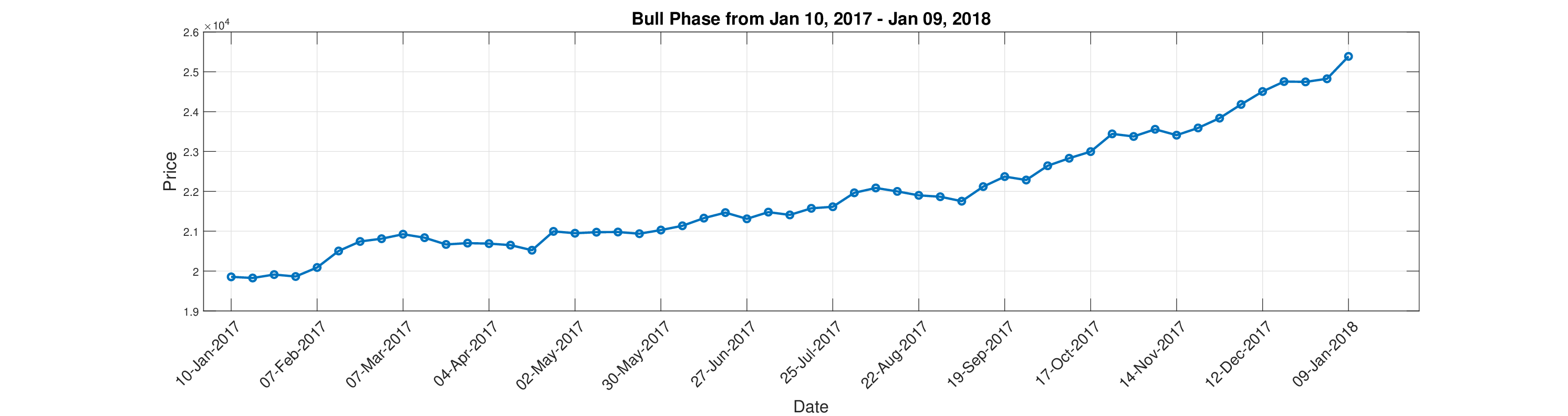}
    \caption{Bullish trend.}
    \label{fig:phasebull}
\end{figure}

In Table \ref{table:bear}, we report the out-of-sample metrics of portfolios from (NoM) and (RoM-RKHS) models in the bearish phase of the DJIA. 
Across all models, the NoMP with $A $=15 has negative mean returns; the RoMP with $A$ = 3 achieves a superior mean return. The NoMP shows lower volatility at $A$ = 3, but its risk increases significantly with increasing $A$. In contrast, the RoMP displays relatively more stable volatility through SD. The NoMP exhibits significantly higher risk in terms of VaR and CVaR. By comparison, RoMP delivers lower VaR and CVaR, particularly when $A$ = 9, indicating better risk mitigation. The NoMP yields positive Jensen’s alpha, but the RoMP significantly outperforms this metric, particularly with $A$ = 3 reflecting higher returns after accounting for market risk. Also, the RoMP achieves the highest omega ratio for $A$ = 15.

From Fig. \ref{fig:bear},  RoMP with $A$ = 3 gives superior mean return in the bearish phase of the DJIA.

{\footnotesize{
\begin{table}[!ht]
   \setlength{\tabcolsep}{1.5pt}
    \centering
    \scalebox{0.8}{
    \begin{tabular}{|c|c|c|c|c|c|c|c|c|c|c|c|c|}
    \hline
    & \multicolumn{5}{c|}{\textbf{(NoM)}} && \multicolumn{5}{c|}{\textbf{(RoM-RKHS)}}& \\
    \hline
    \textbf{Statistics} & \textbf{3} & \textbf{6} & \textbf{9} & \textbf{15} & \textbf{n} & \textbf{BMP} & \textbf{3} & \textbf{6} & \textbf{9} & \textbf{15} & \textbf{n} & \textbf{EQP} \\
    & \textbf{Assets} & \textbf{Assets} & \textbf{Assets} & \textbf{Assets} & \textbf{Assets} &  & \textbf{Assets} & \textbf{Assets} & \textbf{Assets} & \textbf{Assets} & \textbf{Assets} &  \\
 \hline
\textbf{MAX} & 41.546 & 39.093 & 37.862 & 42.709 & 40.137 & 61.793 & 55.57 & 41.039 & 40.09 & 39.563 & 38.922 & \textbf{63.2} \\
\hline
\textbf{MIN} & -96.529 & -114.68 & -114.27 & -109.97 & -114.68 & -129.35 & \textbf{-90.924} & -106.07 & -106.4 & -110.32 & -110.32 & -144.6 \\
\hline
\textbf{MEAN} & -3.39 & -3.805 & -3.554 & -4.478 & -3.608 & -7.822 & \textbf{-1.713} & -1.936 & -2.598 & -3.929 & -3.93 & -6.315 \\
\hline
\textbf{MEDIAN} & -0.033 & -1.499 & -1.162 & -3.01 & -0.838 & -7.053 & 1.785 & 2.707 & \textbf{2.857} & 0.634 & 0.829 & -5.6 \\
\hline
\textbf{SD} & \textbf{23.964} & 25.562 & 25.11 & 25.195 & 25.593 & 31.903 & 26.566 & 25.022 & 26.125 & 26.437 & 26.274 & 35.272 \\
\hline
\textbf{VAR 0.05} & 52.436 & 49.042 & 42.675 & 47.556 & 46.753 & 59.491 & 49.089 & \textbf{43.938} & 54.862 & 53.572 & 53.371 & 69.44 \\
\hline
\textbf{CVAR 0.05} & 75.438 & 78.545 & 77.842 & 74.861 & 79.28 & 96.318 & 78.325 & \textbf{71.449} & 74.523 & 78.248 & 76.708 & 111.07 \\
\hline

\textbf{JENSEN} & 1.101 & 1.131 & 1.388 & 0.778 & 1.362 & 0 & \textbf{3.163} & 2.67 & 2.495 & 1.512 & 1.442 & 2.01 \\
\hline
\textbf{OMEGA} & 674.35 & 741.87 & 799.55 & 817.65 & 753.78 & 768.29 & 584.25 & 642.3 & 683.28 & 626.04 & 555.63 & \textbf{909.29} \\

\hline

\textbf{VAR 0.1} & \textbf{30.969} & 33.746 & 36.652 & 39.863 & 33.311 & 40.815 & 37.52 & 33.542 & 34.64 & 38.354 & 38.915 & 42.69 \\
\hline
\textbf{CVAR 0.1} & 54.215 & 57.345 & 55.714 & 57.572 & 57.491 & 71.136 & 58.265 & \textbf{51.674} & 56.543 & 58.788 & 58.644 & 78.629 \\
\hline
    \end{tabular}}
    \vspace*{0.3cm}
    \caption{The out-of-sample statistics (* $10^{-3}$) for Bearish Phase in the DJIA index.}\label{table:bear}
    
\end{table}
}}

\begin{figure}[htbp]
    \centering
    \includegraphics[width=1\textwidth]{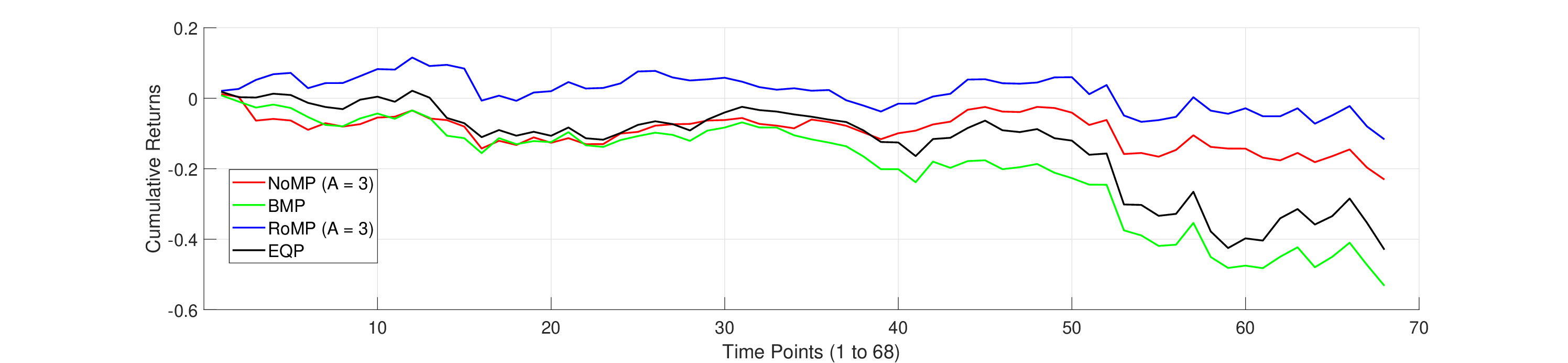} 
    \caption{Cumulative Returns from Portfolios in the Bearish Phase of DJIA}
    \label{fig:bear}
\end{figure}

\newpage
\section{Conclusions}\label{section:6}
In this study, we develop a robust portfolio optimization model designed to minimize mixed conditional value-at-risk (MCVaR) under a chance constraint on portfolio returns, asset investment limits, and a limit on the number of assets. The robustness is integrated by a kernel-based uncertainty set derived from reproducing kernel Hilbert space (RKHS). This facilitates a more accurate depiction of return uncertainty without imposing its specific distribution function. This simplification yielded a second-order cone program (SOCP) formulation for the robust MCVaR model. Testing across six distinct market datasets demonstrated that our model significantly outperformed nominal models, naive portfolios, and benchmark market portfolios. The robust models often produced higher returns with reduced risk. Moreover, the robust models excelled in risk-adjusted performance metrics, such as the Sharpe and Sortino ratios, especially in bearish market conditions, providing better risk mitigation than nominal models, naive portfolios, and benchmark portfolios. This indicates that the proposed models are particularly beneficial in uncertain or volatile market environments, serving as a valuable resource for investors aiming to minimize risk while attaining the desired return. However, we need to conduct further in-depth analysis to see the performance of robust portfolio models in market phases and integrate investors' investment behaviour.

A natural extension of this research would be considering a multi-period portfolio optimization model that accommodates the evolving market conditions over time. Such an enhancement would enable the model to adapt to shifting risk-return profiles, changing economic scenarios, and rebalancing investment decisions. Additionally, integrating practical constraints such as transaction costs, liquidity concerns, investor behavioural cognition, and other real complexities could substantially enhance the model's robustness and applicability in practical scenarios.  
\section*{Data availability}
The data sources have been shared in the article.

\bibliographystyle{elsarticle-num}
\bibliography{sample}

\end{document}